\documentclass[twoside,11pt]{article}

\usepackage{jmlr2e}
\usepackage{amsmath}
\usepackage{amsfonts}
\usepackage{bbm}
\usepackage{booktabs}
\usepackage[ruled,norelsize,vlined]{algorithm2e}
\RequirePackage[colorlinks,citecolor=blue,urlcolor=blue]{hyperref}

\newcommand{\bx}{ {\bf x} }

\newcommand{\by}{ {\bf y} }
\newcommand{\bz}{ {\bf z} }
\newcommand{\bs}{ {\bf s} }
\newcommand{\bX}{ {\bf X} }
\newcommand{\bpi}{ {\boldsymbol \pi} }
\newcommand{\bbeta}{ {\boldsymbol \beta} }
\newcommand{\boeta}{ {\boldsymbol \eta} }
\newcommand{\beps}{ {\boldsymbol \varepsilon} }
\newcommand{\bzero}{ {\bf 0} }
\newcommand{\bv}{ {\bf v} }
\newcommand{\bV}{ {\bf V} }
\newcommand{\bS}{ {\bf S} }
\newcommand{\bU}{ {\bf U} }
\newcommand{\bSigma}{ {\boldsymbol \Sigma} }
\newcommand{\bLambda}{ {\boldsymbol \Lambda} }
\newcommand{\bI}{ {\bf I} }
\newcommand{\bxi}{ {\boldsymbol \xi} }
\newcommand{\bOmega}{ {\boldsymbol \Omega} }
\newcommand{\bDelta}{ {\boldsymbol \Delta} }
\newcommand{\bGamma}{ {\boldsymbol \Gamma} }
\newcommand{\bgamma}{ {\boldsymbol \gamma} }
\newcommand{\bomega}{ {\boldsymbol \omega} }
\newcommand{\post}{ \mbox{\normalfont \tiny pst}}
\newcommand{\new}{ \mbox{\normalfont \scriptsize new}}
\newcommand{\bw}{ {\bf w} }
\newcommand{\bW}{ {\bf W} }
\newcommand{\pr}{ \mbox{pr} }

\newtheorem{Proposition}{Proposition}
\newtheorem{Theorem}{Theorem}
\newtheorem{Corollary}{Corollary}
\newtheorem{Remark}{Remark}
\usepackage{lastpage}
\ShortHeadings{Conjugate Priors for Multinomial Probit Models}{Fasano and Durante}

\firstpageno{1}

\begin{document}

\title{A Class of Conjugate Priors for Multinomial Probit Models which Includes the Multivariate Normal One}

\author{\name Augusto Fasano \email augusto.fasano@unito.it\\
      \addr Department of Economics and Statistics \\
       University of Torino and Collegio Carlo Alberto\\
       Corso Unione Sovietica 218/bis, 10134, Torino, Italy
       \AND
\name Daniele Durante \email daniele.durante@unibocconi.it\\
      \addr Department of Decision Sciences\\ 
      Bocconi Institute for Data Science and Analytics\\
       Bocconi University\\
       Via R\"ontgen 1, 20131, Milan, Italy
}

\editor{}

\vspace{5pt}
\maketitle

\begin{abstract}%   <- trailing '%' for backward compatibility of .sty file
Multinomial probit models are routinely-implemented representations for learning how  the class probabilities of categorical response data change with $p$ observed predictors. Although several frequentist methods have been developed for estimation, inference and classification within such a class of models, Bayesian inference is still lagging behind. This is due to the apparent absence of a tractable class of conjugate priors, that may facilitate posterior inference on the multinomial probit coefficients. Such an issue has motivated increasing efforts toward the development of effective Markov chain Monte Carlo methods, but state-of-the-art solutions still face severe computational bottlenecks, especially in high dimensions. In this article, we show that the entire class of unified skew-normal  (\textsc{sun}) distributions is conjugate to several multinomial probit models. Leveraging this result and the \textsc{sun} properties, we  improve upon state-of-the-art solutions for posterior inference and classification both in terms of closed-form results for several functionals of interest, and also  by developing novel computational methods relying either on independent and identically distributed samples from the exact posterior or on scalable and accurate variational approximations based on blocked partially-factorized  representations. As illustrated in simulations and in a gastrointestinal lesions application, the magnitude of the improvements relative to current methods is particularly evident, in practice, when the focus is on high-dimensional studies.  
\end{abstract}


\vspace{15pt}

\section{Introduction}
\label{sec.1}
Regression models for categorical data are ubiquitous in various fields of application and play a fundamental role in classification \citep[e.g.,][]{agrest_2013}. Within this framework, the overarching goal is to learn how a vector of $L$ class probabilities $\bpi(\bx_i)=[\pi_1(\bx_i), \ldots, \pi_L(\bx_i)]^{\intercal}=[\pr(y_i=1 \mid \bbeta, \bx_i), \ldots, \pr(y_i=L \mid \bbeta,\bx_i)]^{\intercal}$ changes with a set  of $p$ predictors $\bx_i$,   observed for every unit $i=1, \ldots, n$, where $\bbeta$ denotes a vector of coefficients controlling the predictors' effects. We refer to \citet{maddala1986, greene2003} and \citet{agrest_2013} for a broad overview of popular formulations to address such a goal, and focus in this article on the class of multinomial probit models. Indeed, such a broad set of formulations has gained vast popularity in social science, economics and machine learning applications, among others, due to their natural connection with Gaussian regression models, that act as latent predictor-dependent random utilities in a discrete choice setting and also ensure improved interpretability \citep{hausman1978,daganzo2014}. Moreover, expressing predictor-dependent class probabilities via correlated Gaussian latent utilities facilitates improved flexibility, thus avoiding restrictive assumptions, such as the {\em independence of  irrelevant alternatives} \citep{hausman1978}. These desirable properties have stimulated extensive implementations also in the machine learning context \citep[e.g.,][]{girolami2006,rogers2007,riihimaki2013,johndrow2013,agarwal2014, kindo2016}, while motivating several generalizations which extend the classical formulation in \citet{hausman1978} to incorporate class-specific predictor effects \citep{stern_1992} and sequential discrete choices \citep{tutz1991}. 

The aforementioned benefits come, however, with computational difficulties in dealing with integrals of multivariate Gaussian densities \citep[e.g.,][]{genz_1992,horrace2005,chop_2011,botev_2017,genton2018,cao2019,cao2021exploiting}. These  challenges have stimulated an intensive research both in frequentist and in Bayesian settings. In this article, we aim to provide theoretical, methodological and computational advances for the second class of approaches to inference. Indeed, while the frequentist methods for estimation, inference and classification in multinomial probit models are relatively well-established \citep{mcfadden_1989,stern_1992,borsch_1993,geweke1994,natarajan2000monte}, state-of-the-art Bayesian solutions rely either on Markov chain Monte Carlo (\textsc{mcmc}) methods   \citep{Albert_1993, mcculloch1994, nobile1998, mcculloch2000, albert2001, chen2002, imai2005, zhang2006, chan2009,burgette2012, johndrow2013}  or on approximations of the posterior \citep{girolami2006,girolami2007,riihimaki2013,knowles2011}. Despite being widely implemented, both solutions still raise open questions in terms of accuracy, efficiency and computational tractability, especially in large $p$ settings and in imbalanced situations where some classes are relatively less frequent than others. Recalling \citet{chopin_2017,Johndrow2018,Durante2018} and \citet{fasano2019asymptotically}, these issues arise also in simple univariate probit models, and, as discussed in Section~\ref{sec.2}, are even more common in multinomial settings since the dimension of the parameters' space often grows also with the number of classes $L$, due to the inclusion of class-specific effects \citep[e.g.,][]{stern_1992,tutz1991}. In addition, \textsc{mcmc} and approximate methods are still sub-optimal relative to situations in which the posterior is analytically available from a tractable class of distributions.

In Sections \ref{sec.2} and \ref{sec.3}, we generalize recent findings on univariate binary probit regression in \citet{Durante2018}   to prove that the entire class of unified skew-normal  (\textsc{sun}) distributions \citep{arellano_2006}---which includes the classical Gaussian ones as a special case---is a conjugate prior for $\bbeta$ in common multinomial probit models \citep{hausman1978,stern_1992,tutz1991}. Such a general class of distributions has been originally developed in seemingly unrelated contexts to introduce skewness in a multivariate Gaussian density through the cumulative distribution function of another Gaussian vector, thereby retaining several probabilistic properties of multivariate Gaussian variables  \citep{arellano_2006,azzalini_2013}. Leveraging such properties, we derive in Section \ref{sec.3} closed-form expressions for posterior predictive distributions and marginal likelihoods which facilitate classification, model selection and inference, also for  the  parameters regulating the dependence structure among the $L$ alternatives. In fact, although the overarching focus of this article is to provide novel results that facilitate Bayesian inference for the  $\bbeta$ coefficients in multinomial probit models, the closed-form expression we derive for the marginal likelihood is also useful to develop improved methods for point estimation and full Bayesian inference  \citep[][]{mcculloch1994, nobile1998, mcculloch2000,imai2005,chan2009} also on the dependence structure between the different classes; see Section~\ref{sec.3.1} for additional discussion and details.

The evaluation of more complex functionals of the posterior distribution for $\bbeta$ proceeds instead via improved Monte Carlo methods which, unlike for state-of-the-art \textsc{mcmc} routines, rely on independent and identically distributed samples from the exact \textsc{sun} posterior, thus avoid mixing issues and convergence diagnostics. As discussed in Section \ref{sec.3.2.1}, such an improved strategy  deals with multivariate truncated normals and cumulative distribution functions of multivariate Gaussians whose dimension grows with the sample size $n$. Hence, the proposed strategy is particularly useful, in practice, in small-to-moderate $n$ settings, and massively improves state-of-the-art solutions in large $p$ studies, a situation  which occurs in various applications but is computationally impractical under the available implementations \citep{chopin_2017}. To address the scalability issues of the methods proposed  in Section \ref{sec.3.2.1}, we further improve and extend in Section \ref{sec.3.2.2} recent partially-factorized variational methods for univariate probit models   \citep{fasano2019asymptotically} to devise novel blocked partially-factorized  approximations of the posterior distribution in multinomial probit regression which easily scale to large $p$ and $n$ datasets, and almost perfectly match the exact posterior, especially when $p>n$. These findings are further illustrated in a simulation study in Section \ref{sec.4}, and in a gastrointestinal lesions application \citep{mesejo2016} in Section~\ref{sec.5}. Section \ref{sec.6} presents future directions  of research which highlight how these novel results can motivate applied, methodological and computational advances in multinomial probit models. All proofs can be found in Appendix A, and extend conjugacy properties of Gaussian and \textsc{sun}  distributions in probit settings. Initial results on these properties are presented in \citet{Durante2018}, with a focus on Bayesian univariate binary probit regression. These results are a special case of our broader derivations which require  novel extensions to incorporate classical multinomial probit models  \citep{hausman1978}, and related  generalizations \citep{stern_1992,tutz1991}. As clarified in Section \ref{sec.2}, these formulations rely on more complex latent variable representations, typically based on the maximum of a multivariate vector of latent utilities that usually require a separate  treatment relative to the univariate case.

\section{Multinomial Probit Models}
\label{sec.2}
In this section we review three widely-implemented multinomial probit models that cover a large range of applications. These include the classical formulation presented in \citet{hausman1978}, and two subsequent generalizations which account for class-specific predictor effects \citep[][]{stern_1992} and sequential discrete choices \citep[][]{tutz1991}. Despite providing different generative mechanisms for the class probability vector $\bpi(\bx_i)=[\pi_1(\bx_i), \ldots, \pi_L(\bx_i)]^{\intercal}$, all these representations rely on latent Gaussian random utilities and the associated likelihood can be expressed via the cumulative distribution function of a multivariate Gaussian; see~Sections \ref{sec.2.1}--\ref{sec.2.3}. This facilitates the derivation of the conjugacy results for $\bbeta$ in Section \ref{sec.3}. As mentioned in Section \ref{sec.1}, estimation and inference for the parameters quantifying the dependence structure among the class-specific latent utilities is often of interest  \citep[e.g.,][]{mcculloch1994, nobile1998, mcculloch2000,imai2005,chan2009}. Although this goal goes beyond the scope of our contribution, in Section \ref{sec.3.1} we also discuss how the conjugacy results derived for $\bbeta$ can have direct consequences in improving estimation and inference on the dependence among the $L$ alternatives.

\subsection{Classical Discrete Choice  Multinomial  Probit Models}
\label{sec.2.1}
Let us first focus on the classical discrete choice model as originally formulated by  \citet{hausman1978}. Recalling  \cite{greene2003}, this representation expresses each class probability $\pi_l(\bx_i)$ via a random utility model in which every unit $i$ chooses among~$L$ alternatives by maximizing a set of latent Gaussian utilities $z_{i1}, \ldots, z_{iL}$ that depend on $p$-dimensional vectors  of class-specific attributes $\bx_{i1}, \ldots, \bx_{iL}$---encoded in $\bx_i$---as perceived by  unit $i$. More specifically, each class probability $\pi_l(\bx_i)$ in $\bpi(\bx_i)=[\pi_1(\bx_i), \ldots, \pi_L(\bx_i)]^{\intercal}$ can be written as
\begin{eqnarray}
\pr(y_i=l \mid \bbeta,\bx_i)=\pr(z_{il}>z_{ik}, \forall k \neq l)=\pr(\bx^{\intercal}_{il}\bbeta+ \varepsilon_{il}>\bx^{\intercal}_{ik}\bbeta+ \varepsilon_{ik}, \forall k \neq l), 
\label{eq1}
\end{eqnarray}
for every $l=1, \ldots, L$, where $\beps_i =( \varepsilon_{i1}, \ldots,  \varepsilon_{iL})^\intercal \sim \mbox{N}_L(\bzero, \bSigma)$, independently for $i=1, \ldots, n$; see \cite{greene2003} for identifiability restrictions on the matrix $\bSigma$ regulating the dependence among the $L$ alternatives. 

In \eqref{eq1}, the generic vector $\bx_{il}=(x_{il1}, \ldots, x_{ilp})^{\intercal}$ of predictors has elements $x_{ilj}$ measuring how the $j$th attribute of the $l$th alternative is perceived by unit $i$.  For instance, in political studies \citep[e.g.,][]{dow2004}, each $\bx_{il}$ can include both information on voter $i$ and attributes of candidate $l$ as perceived by voter $i$. Hence, this specification assumes that to each individual $i$ are associated $L$ vectors of $p$ observed predictors whose linear combinations $\bx^{\intercal}_{i1}\bbeta, \ldots, \bx^{\intercal}_{iL}\bbeta$ contribute to defining the  $L$ class-specific latent utilities $z_{i1}, \ldots, z_{iL}$. Each individual $i$ will then choose the alternative with the highest random utility $z_{il}=\bx^{\intercal}_{il}\bbeta+\varepsilon_{il}$, which is defined by a deterministic component $\bx^{\intercal}_{il}\bbeta$ with $\bbeta=(\beta_1, \ldots, \beta_p)^\intercal$, plus a Gaussian noise $ \varepsilon_{il}$. This  term accounts for  deviations from the deterministic part due to potential unobserved attributes and, as stated in Proposition \ref{prop1}, it induces a joint likelihood for the observed response data $\by=(y_1, \ldots, y_n)^\intercal$ that coincides with the cumulative distribution function of an  $[n(L-1)]$-variate Gaussian.

\begin{Proposition}
Let $\bv_l$ denote an $L \times 1$  vector having value $1$ in position $l$ and $0$ elsewhere, for every $l=1, \ldots, L$. Moreover, for every $l=1, \ldots, L$, denote with $\bV_{[-l]}$ and $\bX_{i[-l]}$ the $(L-1) \times L$ and $(L-1) \times p$ matrices whose rows are obtained by stacking vectors $(\bv_k-\bv_l)^{\intercal}$ and $(\bx_{il}-\bx_{ik})^\intercal$, respectively, for all $k \neq l$. Then, under the model in \eqref{eq1}, with $\beps_i \sim \mbox{\normalfont N}_L(\bzero, \bSigma)$ independently for every unit $i=1, \ldots, n$, we have
\begin{eqnarray}
p(\by \mid \bbeta, \bX)=\prod_{i=1}^n p(y_i \mid \bbeta, \bx_i)=\prod_{i=1}^n \Phi_{L-1}(\bX_{i[-y_i]} \bbeta; \bV_{[-y_i]} \bSigma \bV^{\intercal}_{[-y_i]})= \Phi_{n(L-1)}(\bar{\bX} \bbeta; \bLambda),
\label{eq2}
\end{eqnarray}
where $\bar{\bX}$ is an $[n(L-1)] \times p$ block matrix with $(L-1) \times p$ row blocks $\bar{\bX}_{[i]}=\bX_{i[-y_i]}$, for each $i=1, \ldots, n$, whereas $\bLambda$ denotes an $[n(L-1)] \times [n(L-1)]$ block diagonal covariance matrix with $(L-1) \times (L-1)$ diagonal blocks $\bLambda_{[ii]}=\bV_{[-y_i]} \bSigma \bV^{\intercal}_{[-y_i]}$, for every $i=1, \ldots, n$. In \eqref{eq2}, the generic function $\Phi_{c}(\bw; \bS)$ denotes the cumulative distribution function, evaluated at $\bw$, of a $c$-variate Gaussian with mean vector $\bzero$ and covariance matrix $\bS$.
\label{prop1}
\end{Proposition}
The above results follow directly from  \eqref{eq1} after noting that $\pr(y_i=l \mid \bbeta,\bx_i)$ can be written as $\pr(\varepsilon_{ik}{-}\varepsilon_{il} < (\bx_{il}{-}\bx_{ik})^\intercal \bbeta, \forall k \neq l) = \pr(\bV_{[-l]}\beps_i< \bX_{i[-l]} \bbeta)=\Phi_{L-1}(\bX_{i[-l]} \bbeta; \bV_{[-l]} \bSigma \bV^{\intercal}_{[-l]}),$
where $\beps_i \sim \mbox{\normalfont N}_L(\bzero, \bSigma)$ and, hence, $\bV_{[-l]}\beps_i  \sim \mbox{\normalfont N}_{L-1}(\bzero, \bV_{[-l]} \bSigma \bV^{\intercal}_{[-l]})$. The final equality in \eqref{eq2} is instead a direct consequence of the properties of multivariate Gaussian random variables. Indeed, since $\bLambda$ is a block diagonal covariance matrix and $\bar{\bX} \bbeta$ is obtained by stacking sub-vectors $\bX_{i[-y_i]} \bbeta$ for $i=1, \ldots, n$, it  follows that $\Phi_{n(L-1)}(\bar{\bX} \bbeta; \bLambda)$ factorizes as the product of $n$ cumulative distribution functions of $(L-1)$-variate Gaussians.

As mentioned previously, this formulation has been originally developed in social science and economic studies where there is a vector of predictors $\bx_{il}$ for each combination of unit $i$ and class $l$ \citep{hausman1978}. This is, however, not always the case in general classification settings. Indeed, in these situations it is more common to observe only a single vector $\bx_i=(x_{i1}, \ldots, x_{ip})^\intercal$ of $p$ predictors for each statistical unit $i=1, \ldots, n$ and the focus is on modeling the vector $\bpi(\bx_i)=[\pi_1(\bx_i), \ldots, \pi_L(\bx_i)]^{\intercal}$, to ultimately predict the class $y_i$ of unit $i$. In Sections \ref{sec.2.2} and \ref{sec.2.3} we focus on two widely-implemented representations \citep{stern_1992,tutz1991}, which address this goal, while still relying on Gaussian latent utilities.

\subsection{Discrete Choice Multinomial Probit Models with Class-Specific Effects}
\label{sec.2.2}
When a single vector $\bx_i=(x_{i1}, \ldots, x_{ip})^\intercal$ of $p$ covariates is observed for each unit $i=1, \ldots, n$, an interpretable and common solution to model differences in the class probabilities within  $\bpi(\bx_i)=[\pi_1(\bx_i), \ldots, \pi_L(\bx_i)]^{\intercal}$ is to introduce class-specific predictors' effects $\bbeta_1, \ldots, \bbeta_L$ as in \citet{stern_1992}, and define again $\pi_l(\bx_i)$ as a function of Gaussian utilities $z_{i1}, \ldots, z_{iL}$ via
\begin{eqnarray}
\pr(y_i=l \mid \bbeta,\bx_i)=\pr(z_{il}>z_{ik}, \forall k \neq l)=\pr(\bx^{\intercal}_{i}\bbeta_l+ \varepsilon_{il}>\bx^{\intercal}_{i}\bbeta_k+ \varepsilon_{ik}, \forall  k \neq l), 
\label{eq3}
\end{eqnarray}
for each $l=1, \ldots, L$, where $\beps_i =( \varepsilon_{i1}, \ldots,  \varepsilon_{iL})^\intercal \sim \mbox{N}_L(\bzero, \bSigma)$, independently for every unit $i=1, \ldots, n$, and $\bbeta_L={\bf 0}$ for identifiability purposes \citep{johndrow2013}. Representation~\eqref{eq3} and its interpretation are closely related to the classical discrete choice multinomial probit model in Section \ref{sec.2.1}, with the only exception that the differences in the class-specific latent utilities $z_{i1}, \ldots, z_{iL}$, $i=1, \ldots, n$, are now driven by changes in the  vectors of coefficients   $\bbeta_1, \ldots, \bbeta_L$,  rather than in the vectors of predictors as in model \eqref{eq1}. For instance, recalling the political example discussed in Section \ref{sec.2.1}, although the age is an attribute specific to voter $i$, it is reasonable to expect that such a covariate has a different effect in producing the utilities $z_{i1}=\bx^{\intercal}_{i}\bbeta_1+ \varepsilon_{i1}, \ldots, z_{iL}=\bx^{\intercal}_{i}\bbeta_L+ \varepsilon_{iL}$ that voter $i$ assigns to the different candidates $l=1, \ldots, L$. This property can be included by allowing the coefficient associated with the age attribute to change across classes,  thus providing a formulation more similar to classical multinomial logit models \citep[e.g.,][]{greene2003}, relative to \eqref{eq1}. As stated in Proposition \ref{prop2}, also under this representation the likelihood for  the observed response data $\by=(y_1, \ldots, y_n)^\intercal$ coincides with the cumulative distribution function of an  $[n(L-1)]$-variate Gaussian.

\begin{Proposition}
Denote with $\bv_l$ the $L \times 1$ vector with value $1$ in position $l$ and $0$ elsewhere, for each $l=1, \ldots, L$. Moreover, let $\bx_{il}=\bar{\bv}_l \otimes \bx_i$, where $\bar{\bv}_l$ is the $(L-1) \times 1$ vector obtained by the removing the $L$-th element from $\bv_l$, whereas $\otimes$ denotes the Kronecker product. Then, under model \eqref{eq3} with $\beps_i \sim \mbox{\normalfont N}_L(\bzero, \bSigma)$, independently for each unit $i=1, \ldots, n$, we have
\begin{eqnarray}
p(\by \mid \bbeta, \bX)=\prod_{i=1}^n p(y_i \mid \bbeta, \bx_i)=\prod_{i=1}^n \Phi_{L-1}(\bX_{i[-y_i]} \bbeta; \bV_{[-y_i]} \bSigma \bV^{\intercal}_{[-y_i]})= \Phi_{n(L-1)}(\bar{\bX} \bbeta; \bLambda),
\label{eq4}
\end{eqnarray}
where $\bbeta=(\bbeta^{\intercal}_1, \ldots, \bbeta^{\intercal}_{L-1})^{\intercal}$, while $\bX_{i[-y_i]}$, $\bV_{[-y_i]}$, $\bar{\bX}$ and $\bLambda$ are defined as in Proposition~\ref{prop1}, after setting $\bx_{il}=\bar{\bv}_l\otimes\bx_i$ for each $i=1, \ldots, n$ and $l=1, \ldots, L$. Hence, in this case $\bX_{i[-y_i]}$ and $\bar{\bX}$ have dimension $(L-1) \times [p(L-1)]$ and $[n(L-1)] \times [p(L-1)]$, respectively. 
\label{prop2}
\end{Proposition}
Proposition \ref{prop2} follows as a directed consequence of Proposition \ref{prop1}, upon noticing that model \eqref{eq3} can be re-written as a particular case of model \eqref{eq1} with working covariates $\bx_{il}$ as defined in Proposition \ref{prop2}. Indeed, note that by setting $\bx_{il}=\bar{\bv}_l\otimes\bx_i$, $i=1, \ldots, n$, $l=1, \ldots, L$ and $\bbeta=(\bbeta^{\intercal}_1, \ldots, \bbeta^{\intercal}_{L-1})^{\intercal}$, the class probabilities in \eqref{eq3} can be re-expressed as $\pr(y_i=l \mid \bbeta,\bx_i)=\pr(z_{il}>z_{ik}, \forall k \neq l)=\pr(\bx^{\intercal}_{i}\bbeta_l+ \varepsilon_{il}>\bx^{\intercal}_{i}\bbeta_k+ \varepsilon_{ik}, \forall  k \neq l)=\pr(\bx^{\intercal}_{il}\bbeta+ \varepsilon_{il}>\bx^{\intercal}_{ik}\bbeta+ \varepsilon_{ik}, \forall  k \neq l)$, for $l=1, \ldots, L$,  where the last quantity is the equation for the class probabilities in  \eqref{eq1}.

\subsection{Sequential Discrete Choice  Multinomial  Probit Models}
\label{sec.2.3}
Before focusing on prior specification and posterior derivations, we consider also  an extension of the sequential discrete choice  multinomial probit model studied in \citet{albert2001} and originally proposed by \citet{tutz1991}. Such a model still relies on a set of class-specific latent utilities but  is conceptually different from those presented in Sections \ref{sec.2.1} and  \ref{sec.2.2}, since the choice among the $L$ classes is modeled via a nested sequence of binary decisions where the generic step $l$ of this sequential decision process is reached if individual $i$ has not chosen classes $1, \ldots, l-1$. At this step, the binary decision  will be to either pick class $l$ with probability $\pr(y_i=l \mid y_i > l-1, \bbeta, \bx_i)=\Phi(\bx^{\intercal}_{i}\bbeta_l)$ or to consider one of the subsequent alternatives $l+1, \ldots, L$ with complement probability $\pr(y_i>l \mid y_i > l-1, \bbeta, \bx_i)=1-\Phi(\bx^{\intercal}_{i}\bbeta_l)$. Note that relative to the original formulations in \citet{albert2001} and \citet{tutz1991}, here we consider a slightly different reparameterization and also allow the entire vector of coefficients, and not just the intercept, to change with the different labels, thus providing a more general representation. As discussed by \citet{albert2001} also this model has a latent utility representation which expresses each $\pi_l(\bx_i)$ in $\bpi(\bx_i)=[\pi_1(\bx_i), \ldots, \pi_L(\bx_i)]^{\intercal}$ as
\begin{eqnarray}
\pr(y_i=l \mid \bbeta,\bx_i)=\pr(z_{il}>0)\prod_{k=1}^{l-1}p(z_{ik}<0)=\pr(\bx^{\intercal}_{i}\bbeta_l{+} \varepsilon_{il}>0) \prod_{k=1}^{l-1}\pr(\bx^{\intercal}_{i}\bbeta_k{+} \varepsilon_{ik}<0), 
\label{eq5}
\end{eqnarray}
for $l=1, \ldots, L-1$, and $\pr(y_i=L \mid \bbeta,\bx_i)=\prod\nolimits_{k=1}^{L-1}\pr(\bx^{\intercal}_{i}\bbeta_k+ \varepsilon_{ik}<0)$, where $\varepsilon_{il} \sim \mbox{N}(0,1)$ independently for every unit $i=1, \ldots, n$ and class $l=1, \ldots, L-1$. Model \eqref{eq5} provides a general representation in which each $z_{il}=\bx^{\intercal}_{i}\bbeta_l+ \varepsilon_{il}$ is the utility of choosing alternative $l$ against the subsequent ones $l+1, \ldots, L$, given that the classes $1, \ldots, l-1$ have not been selected in the previous steps of the sequential decision process. Proposition \ref{prop3} shows that, although conceptually different from the models in Sections \ref{sec.2.1}--\ref{sec.2.2}, also such a formulation admits a similar expression for the joint  likelihood of the data $\by=(y_1, \ldots, y_n)^{\intercal}$.

\begin{Proposition}
Define $\bar{\by}_i=({\bf 0}^{\intercal}_{y_i-1},1)^{\intercal}$ if $y_i \leq L-1$, and $\bar{\by}_i={\bf 0}_{L-1}$ if $y_i=L$, where the generic ${\bf 0}_{c}$ is a $c \times 1$ vector of zeroes. Moreover, let $\bar{n}=n_1+ \cdots + n_n$ with $n_i=\min(y_i,L-1)$. Then, under \eqref{eq5} with $\varepsilon_{il} \sim \mbox{\normalfont N}(0,1)$ independently for $i=1, \ldots, n$, $l=1, \ldots, L-1$, we have
\begin{eqnarray}
p(\by \mid \bbeta, \bX)=\prod_{i=1}^n p(y_i \mid \bbeta, \bx_i)=\prod_{i=1}^n \Phi_{n_i}(\bX_{i} \bbeta; \bI_{n_i})= \Phi_{\bar{n}}(\bar{\bX} \bbeta; \bLambda),
\label{eq6}
\end{eqnarray}
where  $\bbeta=(\bbeta^{\intercal}_1, \ldots, \bbeta^{\intercal}_{L-1})^{\intercal}$, $\bLambda=\bI_{\bar{n}}$ and $\bar{\bX}$ denotes an $\bar{n} \times [p(L-1)]$ matrix with $n_i \times  [p(L-1)]$ row blocks $\bar{\bX}_{[i]}=\bX_i$ defined as $\bX_i=(\mbox{\normalfont diag}(2\bar{\by}_i- {\bf 1}) \otimes \bx_i^{\intercal},\bzero_{n_i \times [p(L-1-n_i)]})$, for every statistical unit  $i=1, \ldots, n$. In \eqref{eq6}, the quantity $\bI_{n_i}$ refers to the $n_i \times n_i$ identity matrix.
\label{prop3}
\end{Proposition}
To clarify Proposition \ref{prop3}, it suffices to re-write $\pr(y_i=l \mid \bbeta,\bx_i)$, $l=1, \ldots,L-1$, in \eqref{eq5}, as $ \Phi(\bx^{\intercal}_{i}\bbeta_l) \prod\nolimits_{k=1}^{l-1}[1-\Phi(\bx^{\intercal}_{i}\bbeta_k)]=\prod\nolimits_{k=1}^{l}\Phi[(2\bar{y}_{ik}-1)\bx^{\intercal}_{i}\bbeta_k]=\Phi_{l}(\bX_{i} \bbeta; \bI_{l}),$ where $\bar{\by}_i$ is defined as in Proposition \ref{prop3}. The above result leverages standard properties of  multivariate Gaussians.

Combining Propositions \ref{prop1}--\ref{prop3} it is clear that, despite characterizing different utility-based decision mechanisms, models \eqref{eq1}, \eqref{eq3} and \eqref{eq5} have a similar form for the joint likelihood. The only difference among such likelihoods is the dimension of the  cumulative distribution functions and the definition of the known matrices $\bar{\bX}$ and $\bLambda$, which change depending on the type of model. These results are fundamental for the novel conjugacy results  in Section~\ref{sec.3}.

\section{Conjugate Bayesian Inference for Multinomial Probit Models}
\label{sec.3}
Common Bayesian implementations of multinomial probit models consider a multivariate Gaussian prior $\mbox{N}_q(\bxi, \bOmega)$ for the parameters in $\bbeta$, where $q$ is equal to $p$ in model  \eqref{eq1} and to $p(L-1)$ in models \eqref{eq3} and \eqref{eq5}, whereas $\bxi$ and $\bOmega$ denote the pre-specified prior mean vector and covariance matrix, respectively \citep{Albert_1993, mcculloch1994, nobile1998, mcculloch2000, albert2001, chen2002, imai2005, zhang2006, burgette2012, johndrow2013}. Besides providing a default specification in various Bayesian regression models, this choice is also motivated by the Gaussian assumption for the latent utilities in \eqref{eq1}, \eqref{eq3} and \eqref{eq5} which implies an augmented data representation facilitating the implementation of \textsc{mcmc} \citep[e.g.,][]{Albert_1993,  albert2001, imai2005, holmes_2006,chopin_2017} and approximate  methods \citep[e.g.,][]{girolami2006,girolami2007,riihimaki2013,knowles2011} for inference and prediction.

As discussed in Section \ref{sec.1}, the above strategies have computational drawbacks---especially in large $p$ settings---and are motivated by the apparent absence of conjugacy between multinomial probit likelihoods and the Gaussian prior for $\bbeta$. In Section \ref{sec.3.1}, we show not only that the posterior in this setting is a \textsc{sun}, but also that the whole \textsc{sun} family is conjugate to multinomial probits, thereby obtaining closed-form posterior distributions under a broad variety of priors, which include also the default Gaussian one and, as a byproduct, Gaussian processes. Leveraging the novel results in Section \ref{sec.3.1}, we develop in Section \ref{sec.3.2} improved Monte Carlo methods for full Bayesian inference and classification, along with scalable and accurate approximations of the \textsc{sun} posterior in high-dimensional settings.

Before providing an overview of the \textsc{sun} distribution \citep{arellano_2006,azzalini_2013} and presenting our conjugacy results, we shall emphasize that some of the aforementioned contributions  consider also priors for $\bSigma$ in models \eqref{eq1} and \eqref{eq3}. Recalling Sections \ref{sec.1}--\ref{sec.2}, our focus in this article is on the posterior for $\bbeta$ conditioned on $\bSigma$ and, therefore, we avoid  additional identifiability and computational complications which arise when including a prior also for $\bSigma$. Nonetheless, as discussed in Section \ref{sec.3.1}, the closed-form expression for the marginal likelihood $p(\by \mid \bX)$ presented in Corollary \ref{cor2}, and the i.i.d.\ sampler to generate values from the posterior $p(\bbeta \mid \by, \bX)$ outlined in Algorithm \ref{algo:samplingIID}, can be useful to improve both point estimation and full Bayesian inference also on $\bSigma$.

\subsection{Conjugacy via Unified Skew-Normal Priors}
\label{sec.3.1}
Consistent with Section \ref{sec.3}, let us assume a $\textsc{sun}_{q,h}(\bxi, \bOmega, \bDelta, \bgamma, \bGamma)$ prior for $\bbeta$, whose density
\begin{eqnarray}
p(\bbeta)=\phi_q(\bbeta-\bxi;\bOmega)\frac{\Phi_h(\bgamma+\bDelta^{\intercal}\bar{\bOmega}^{-1}\bomega^{-1}(\bbeta-\bxi);\bGamma-\bDelta^{\intercal}\bar{\bOmega}^{-1}\bDelta)}{\Phi_h(\bgamma;\bGamma)},
\label{eq7}
\end{eqnarray}
is obtained by modifying the density function $\phi_q(\bbeta-\bxi;\bOmega)$ of a $q$-variate Gaussian $\mbox{N}_{q}(\bxi, \bOmega)$, via a skewness-inducing mechanism driven by the cumulative  distribution function, computed at $\bgamma+\bDelta^{\intercal}\bar{\bOmega}\phantom{.}^{-1}\bomega\phantom{.}^{-1}(\bbeta-\bxi) \in \mathbb{R}^h$, of an $h$-variate Gaussian with  mean vector ${\bf 0}$ and  $h \times h$ covariance matrix $\bGamma-\bDelta^{\intercal}\bar{\bOmega}\phantom{.}^{-1}\bDelta$. The quantity $\Phi_h(\bgamma;\bGamma)$ is instead the normalizing constant, which coincides with the cumulative distribution function, evaluated at $\bgamma  \in \mathbb{R}^h$, of an $h$-variate Gaussian with  mean vector $\bzero$ and $h \times h$ covariance matrix $\bGamma$. As is clear from \eqref{eq7}, the amount of skewness in the prior is mainly controlled by the $q \times h$ matrix $\bDelta$. Indeed, when all the entries in $\bDelta$ are $0$, the prior $p(\bbeta)$ in \eqref{eq7} coincides with the density of a $q$-variate Gaussian with mean vector $\bxi$ and  covariance matrix $\bOmega=\bomega\bar{\bOmega} \bomega$ obtained via the quadratic combination among the correlation matrix $\bar{\bOmega}$ and the diagonal scale matrix $\bomega=({\bOmega} \odot \bI_{q})^{1/2} $, where $\odot$ is the element-wise Hadamard product. Such a class of Gaussian priors can be also easily obtained by setting $h=0$ in \eqref{eq7}. As discussed in \citet{arellano_2006}, the multivariate Gaussian case is just an example of a broad variety of distributions which can be obtained from prior \eqref{eq7} under suitable choices for its parameters. Additional priors of interest within this class are independent univariate skew-normals \citep{azza_1985} for the coefficients in $\bbeta$ and classical multivariate skew-normals \citep{azza_1996} for the entire vector $\bbeta$. Therefore, our results allow tractable inference in Bayesian multinomial probit models under a broad class of priors that include Gaussian specifications along with asymmetric priors which may be useful in social science and econometric applications. Note that also non-linear effects modeled via Gaussian processes induce a multivariate Gaussian prior and, hence, our results can be directly applied to  the flexible classification strategies discussed in \citet{girolami2006} and \citet{riihimaki2013}, among others.

To further clarify the main roles of the parameters $\bxi, \bOmega, \bDelta, \bgamma$ and $\bGamma$ note that, as shown in \citet{arellano_2006}, if $\bbeta \sim \textsc{sun}_{q,h}(\bxi, \bOmega, \bDelta, \bgamma, \bGamma)$, then
\begin{eqnarray}
\bbeta\stackrel{\mbox{\scriptsize d}}{=}\bxi+\bomega(\bV_0+\bDelta\bGamma^{-1}\bV_1), \quad \mbox{with} \ \bV_0 \sim\mbox{N}_q(\boldsymbol{0},\bar{\bOmega}-\bDelta\bGamma^{-1}\bDelta^\intercal), \ \bV_1\sim\mbox{TN}_h(-\bgamma;\boldsymbol{0},\bGamma), 
\label{eq8}
\end{eqnarray}
where $\mbox{TN}_h(-\bgamma;\boldsymbol{0},\bGamma)$ denotes an $h$-variate Gaussian with zero mean, covariance matrix $\bGamma$ and truncation below $-\bgamma$. Recalling \citet{arellano_2006}, representation~\eqref{eq8} also relates to a simple conditioning mechanism. In particular, if $\bU_1\in \mathbb{R}^h$ and $\bU_0 \in \mathbb{R}^q$ are two random vectors jointly distributed as a $\mbox{N}_{h+q}({\bf 0}, \bOmega^*)$, where $\bOmega^*$ denotes a correlation  matrix with blocks $\bOmega_{[11]}^*=\bGamma$, $\bOmega_{[22]}^*=\bar{\bOmega}$ and $\bOmega_{[21]}^*=(\bOmega^{*}_{[12]})^{\intercal}=\bDelta$, then $\bbeta=\bxi+\bomega\bar{\bbeta}$, with $\bar{\bbeta}=(\bU_0 \mid \bU_1+\bgamma> {\bf 0})$, has \textsc{sun} distribution with density as in \eqref{eq7}. Hence, $\bxi$ and $\bomega$ mostly regulate the location and the scale of the prior, while $\bDelta$, $\bGamma$, and $\bar{\bOmega}$ control dependence and skewness. Finally, $\bgamma$ defines the truncation threshold in the conditioning part.

Besides clarifying the role of the prior parameters, the additive representation \eqref{eq8} of the \textsc{sun} random variable is useful also for posterior inference since, as we will discuss, it provides a direct strategy to sample i.i.d.\ values from the \textsc{sun} distribution, thereby improving upon state-of-the-art \textsc{mcmc} methods for Bayesian multinomial probit models. Indeed, as shown in Theorem \ref{teo1}, the \textsc{sun} prior in \eqref{eq7} is conjugate to the multinomial probit likelihoods reported in \eqref{eq2}, \eqref{eq4} and \eqref{eq6}, meaning that also the posterior $(\bbeta \mid \by, \bX)$ has a \textsc{sun} distribution. In particular $(\bbeta \mid \by, \bX) \sim \textsc{sun}_{q,h+m}(\bxi_{\post}, \bOmega_{\post}, \bDelta_{\post}, \bgamma_{\post}, \bGamma_{\post})$; see Table \ref{t1} for details on the specific dimensions of the \textsc{sun} posterior under the three multinomial probit models discussed in Section~\ref{sec.2}, when considering either $\textsc{sun}_{q,h}(\bxi, \bOmega, \bDelta, \bgamma, \bGamma)$ or $\mbox{N}_q(\bxi,\bOmega)$ priors.

\begin{table}[t]
\centering
\begin{tabular}{cc|ccc} 
 &&\multicolumn{3}{c}{$\textsc{sun}_{q,h+m}(\bxi_{\post}, \bOmega_{\post}, \bDelta_{\post}, \bgamma_{\post}, \bGamma_{\post})$} \\
\hline
Model& Prior & $q$& \qquad$h$&$m$ \\
\hline
Model \eqref{eq1} (Proposition \ref{prop1}) && $p$ & \qquad$h$ & $n(L-1)$ \\ 
Model \eqref{eq3} (Proposition \ref{prop2}) & $\textsc{sun}_{q,h}(\bxi, \bOmega, \bDelta, \bgamma, \bGamma)$ &$p(L-1)$ & \qquad$h$ & $n(L-1)$ \\ 
Model \eqref{eq5} (Proposition \ref{prop3}) & &$p(L-1)$ & \qquad $h$ & $\bar{n}$ \\ 
\hline
Model \eqref{eq1} (Proposition \ref{prop1}) && $p$ & \qquad$0$ & $n(L-1)$ \\ 
Model \eqref{eq3} (Proposition \ref{prop2}) & $\mbox{N}_q(\bxi,\bOmega)$ &$p(L-1)$ &\qquad $0$ & $n(L-1)$ \\ 
Model \eqref{eq5} (Proposition \ref{prop3}) & &$p(L-1)$ &\qquad $0$ & $\bar{n}$ \\ 
\hline
\end{tabular}
\caption{Dimension of the $\textsc{sun}_{q,h+m}(\bxi_{\post}, \bOmega_{\post}, \bDelta_{\post}, \bgamma_{\post}, \bGamma_{\post})$ posterior for $\bbeta$ in the multinomial probit models in Section~\ref{sec.2}, when considering either \textsc{sun} or Gaussian priors.}
\label{t1}
\end{table}

\begin{Theorem}
Let $p(\bbeta)$ denote the \textsc{sun} prior  density in   \eqref{eq7}, and define with $\Phi_m(\bar{\bX} \bbeta; \bLambda)$ the generic multinomial probit likelihood reported in   \eqref{eq2}, \eqref{eq4} and \eqref{eq6}, with $m$, $\bar{\bX}$ and $\bLambda$ defined as in Propositions \ref{prop1}, \ref{prop2} or \ref{prop3} depending on whether the focus is on the model  \eqref{eq1}, \eqref{eq3} or \eqref{eq5}, respectively; see also Table \ref{t1}. Then, the posterior density $p(\bbeta \mid \by, \bX)$ of $\bbeta$ is 
\begin{eqnarray}
p(\bbeta \mid \by, \bX)=\phi_q(\bbeta-\bxi_{\post};\bOmega_{\post})\frac{\Phi_{h{+}m}(\bgamma_{\post}{+}\bDelta_{\post}^{\intercal}\bar{\bOmega}_{\post}^{-1}\bomega_{\post}^{-1}(\bbeta-\bxi_{\post});\bGamma_{\post}{-}\bDelta_{\post}^{\intercal}\bar{\bOmega}_{\post}^{-1}\bDelta_{\post})}{\Phi_{h+m}(\bgamma_{\post};\bGamma_{\post})},
\label{eq9}
\end{eqnarray}
with $\bxi_{\post}=\bxi$, $\bOmega_{\post}=\bOmega$, $\bDelta_{\post}=(\bDelta, \bar{\bOmega}\bomega\bar{\bX}^{\intercal}\bs^{-1})$, $\bgamma_{\post}=(\bgamma^{\intercal},\bxi^{\intercal}\bar{\bX}^{\intercal}\bs^{-1})^{\intercal}$, while $\bGamma_{\post}$ is an $(h+m) \times (h+m)$ covariance matrix with blocks $\bGamma_{\post[11]}=\bGamma$, $\bGamma_{\post[22]}=\bs^{-1}(\bar{\bX}\bOmega\bar{\bX}^{\intercal}+\bLambda)\bs^{-1}$ and $\bGamma_{\post[21]}=\bGamma^{\intercal}{\phantom{}}_{\post[12]}=\bs^{-1}\bar{\bX}\bomega\bDelta$, where $\bs=[(\bar{\bX}\bOmega\bar{\bX}^{\intercal}+\bLambda)\odot \bI_{m}]^{1/2}$. Note that in \eqref{eq9}, the dimension $q$ is equal to $p$ under model \eqref{eq1}, and to $p(L-1)$ under models  \eqref{eq3} and  \eqref{eq5}. 
\label{teo1}
\end{Theorem}

\begin{Remark}
As a  consequence of Theorem \ref{teo1}, it follows that also the multivariate Gaussian prior---which provides a special case of unified skew-normal---yields to a \textsc{sun} posterior when updated with the multinomial probit likelihoods in \eqref{eq2},  \eqref{eq4} and  \eqref{eq6}. In particular, if $p(\bbeta)=\phi_q(\bbeta-\bxi;\bOmega)$ it immediately follows from Theorem \ref{teo1} that the posterior distribution is a \textsc{sun} having density as in \eqref{eq9}, with $h=0$ and posterior parameters $\bxi_{\post}=\bxi$, $\bOmega_{\post}=\bOmega$, $\bDelta_{\post}=\bar{\bOmega}\bomega\bar{\bX}^{\intercal}\bs^{-1}$, $\bgamma_{\post}=\bs^{-1}\bar{\bX}\bxi$, $\bGamma_{\post}=\bs^{-1}(\bar{\bX}\bOmega\bar{\bX}^{\intercal}+\bLambda)\bs^{-1}$, where $\bs=[(\bar{\bX}\bOmega\bar{\bX}^{\intercal}+\bLambda)\odot \bI_{m}]^{1/2}$. 
\end{Remark}

Theorem \ref{teo1} generalizes Corollary 4 in \citet{Durante2018} to provide novel results with important implications in Bayesian inference for multinomial probit models. As discussed in  \citet{arellano_2006} \textsc{sun} distributions share several common properties with multivariate Gaussians. A relevant one is that this family is closed under marginalization, linear combinations and conditioning. Within our context, this means that the posterior for each single coefficient and linear combinations of interest---such as those defining the latent utilities---are still \textsc{sun} and their parameters can be obtained via simple transformations of those in Theorem \ref{teo1} \citep{arellano_2006,azzalini_2013}. According to \eqref{eq9}, also the normalizing constant  of the posterior is available in closed form and coincides with the cumulative distribution function $\Phi_{h+m}(\bgamma_{\post};\bGamma_{\post})$, evaluated at $\bgamma_{\post}$, of a multivariate Gaussian with ${\bf 0}$ mean and covariance matrix $\bGamma_{\post}$. As outlined in Corollaries~\ref{cor2} and \ref{cor3}, this result is fundamental to obtain closed--form expressions for marginal likelihoods and predictive distributions, that are useful for model selection and classification.

\begin{Corollary}
Under the settings in Theorem \ref{teo1}, the marginal likelihood  can be expressed as
\begin{eqnarray}
p(\by \mid \bX)=\frac{p(\by,\bbeta \mid \bX)}{p(\bbeta \mid \by, \bX)}=\frac{\Phi_{h+m}(\bgamma_{\post};\bGamma_{\post})}{\Phi_h(\bgamma;\bGamma)}, 
\label{eq10}
\end{eqnarray}
with $\bgamma_{\post}$ and $\bGamma_{\post}$ defined as in Theorem \ref{teo1}. \label{cor2}
\end{Corollary}

\begin{Corollary}
Consider the expanded dataset in which, besides the original data $\by$ and $\bX$, we also have an additional unit with predictors $\bx_{\new}$ and response $y_{\new}=l$. Moreover, let $m_l$, $\bar{\bX}_l$ and $\bLambda_l$ be defined analogously to $m$, $\bar{\bX}$ and $\bLambda$ in Theorem~\ref{teo1}, when the expanded dataset is considered. Then, under the settings of Theorem \ref{teo1}, we have that 
\begin{eqnarray}
\mbox{\normalfont pr}(y_{\new}=l \mid \by,\bX, \bx_{\new})=\frac{\mbox{\normalfont pr}(y_{\new}=l,\by  \mid \bX,\bx_{\new})}{p(\by \mid \bX,\bx_{\new})}=\frac{\Phi_{h+m_{l}}(\bgamma_{l\post};\bGamma_{l\post})}{\Phi_{h+m}(\bgamma_{\post};\bGamma_{\post})}, 
\label{eq11}
\end{eqnarray}
for each $l=1, \ldots, L$, with $\bgamma_{\post}$ and $\bGamma_{\post}$ as in Theorem \ref{teo1}, while $\bgamma_{l\post}$ and $\bGamma_{l\post}$ coincide with $\bgamma_{\post}$ and $\bGamma_{\post}$, evaluated at $\bar{\bX}_l$ and $\bLambda_l$, instead of $\bar{\bX}$ and $\bLambda$.
\label{cor3}
\end{Corollary}

Corollaries \ref{cor2}--\ref{cor3} facilitate closed-form Bayesian model selection and classification without the need to rely on \textsc{mcmc}. Moreover, although point estimation and full Bayesian inference on $\bSigma$ goes beyond the scope of the present contribution, as anticipated in Sections \ref{sec.1} and~\ref{sec.2}, Corollary \ref{cor2} is practically relevant also for improving current solutions addressing this goal \citep[e.g.,][]{mcculloch1994, nobile1998, mcculloch2000,imai2005,chan2009}. For example, estimation for the parameters in  $\bSigma$ could proceed via direct maximization of the marginal likelihood in~\eqref{eq10}, after integrating out $\bbeta$ analytically. Also full Bayesian inference for $\bSigma$ can benefit from the closed-form marginal likelihood in Corollary \ref{cor2}, since it allows implementation of collapsed Metropolis--Hastings schemes that produce samples from the posterior of  $\bSigma$ after integrating out $\bbeta$ analytically. This strategy is expected to yield gains in mixing relative to common \textsc{mcmc} methods that leverage full-conditional distributions depending both on $\bbeta$ and, potentially, on augmented data \citep{park2009}. Such advantages are practically relevant and notable in settings where $\bbeta$ is high-dimensional and the size of the distribution functions in Corollary~\ref{cor2} is small-to-moderate, thereby allowing accurate and rapid evaluation of  \eqref{eq10} at different $\bSigma$  via recent strategies \citep{botev_2017, genton2018,cao2019,cao2021exploiting}. 

Exploiting the moment generating function of the \textsc{sun}  in Section 2.3 of  \citet{arellano_2006}  and the additional derivations in \citet{azzalini_2010,gupta_2013} and \citet{azzalini_2013} closed-form expressions can be derived also for the posterior mean, covariance matrix and cumulative distribution function of $\bbeta$, thereby facilitating Bayesian point estimation, uncertainty quantification and classification. Such expressions require, however, the evaluation of multivariate Gaussian cumulative distribution functions and  tedious derivations that do not facilitate calculation of more complex functionals, thus motivating the alternative computational methods presented in Section~\ref{sec.3.2}. 

\subsection{Computational Methods}
\label{sec.3.2}
This section provides new computational methods for Bayesian multinomial probit models that exploit results in Section \ref{sec.3.1} to improve upon state-of-the-art routines, especially in large $q$ settings. In particular, in Section \ref{sec.3.2.1} we derive Monte Carlo methods that, unlike current \textsc{mcmc} solutions, rely on independent and identically distributed samples from the exact \textsc{sun} posterior in \eqref{eq9}. Such a strategy requires to sample from $(h+m)$-variate truncated normals with full covariance matrix and, hence, becomes impractical as $h+m$ grows. To address this issue, we also propose in Section \ref{sec.3.2.2} a blocked partially-factorized  variational Bayes that relaxes various factorization assumptions of classical mean-field families to obtain more accurate and computationally efficient approximations, that almost perfectly match the exact posterior in large $q$ settings, especially when $q>h+m$; see Table \ref{t1} for details on how $q$, $h$ and $m$ relate to $p$, $n$ and $L$ under the multinomial probit models  in Sections~\ref{sec.2.1}--\ref{sec.2.3}.

\begin{algorithm*}[t]
	\caption{Strategy to sample  i.i.d.\ from the \textsc{sun} posterior in Theorem \ref{teo1}} 
		\For{$t = 1,\ldots,T$}{
		{\bf [1]} Sample $\bV^{(t)}_0 \sim\mbox{N}_q(\boldsymbol{0},\bar{\bOmega}_{\post}-\bDelta_{\post}\bGamma_{\post}^{-1}\bDelta_{\post}^\intercal)$ [in \texttt{R} use  \texttt{rmvnorm}] \\
		{\bf [2]} Sample  $\bV^{(t)}_1 \sim\mbox{TN}_{h+m}(-\bgamma_{\post};\boldsymbol{0},\bGamma_{\post})$ [in \texttt{R} use \texttt{rtmvnorm}  \citep{botev_2017}]  \\
{\bf [3]} Set $\bbeta^{(t)}=\bxi_{\post}+\bomega_{\post}(\bV^{(t)}_0+\bDelta_{\post}\bGamma_{\post}^{-1}\bV^{(t)}_1)$ \\
	}
	{\bf Output:} i.i.d. samples $\bbeta^{(1)}, \ldots, \bbeta^{(T)}$ from  \eqref{eq9}. Based on such samples, posterior functionals $\mathbb{E}[g(\bbeta) \mid \by, \bX]$ can be computed, via Monte Carlo, as $\sum_{t=1}^T g(\bbeta^{(t)})/T$.	
		\label{algo:samplingIID}
\end{algorithm*}

\subsubsection{Monte Carlo Methods via Independent Samples from the Posterior}
\label{sec.3.2.1}
Complex functionals of the posterior can be effectively evaluated via Monte Carlo methods leveraging the additive representation of the \textsc{sun}  in  \eqref{eq8}. This  allows to sample independent and identically distributed (i.i.d.) values from the posterior  in Theorem \ref{teo1}, via linear combinations among samples from multivariate Gaussians and multivariate truncated normals. As outlined in Algorithm \ref{algo:samplingIID}, this routine crucially avoids \textsc{mcmc} methods, thus circumventing convergence and mixing issues commonly seen in Bayesian multinomial probit  \citep{johndrow2013}, while allowing parallel implementations. A possible computational drawback in Algorithm~\ref{algo:samplingIID} is  sampling from $\mbox{TN}_{h+m}(-\bgamma_{\post};\boldsymbol{0},\bGamma_{\post})$. Recent advances based on minimax tilting methods \citep{botev_2017} have made this task  computationally feasible for multivariate truncated normals with a dimension of few hundreds, thereby making Algorithm \ref{algo:samplingIID} an efficient   strategy in small-to-moderate $h+m$ and large, potentially huge, $q$ studies. Recalling \citet{chopin_2017}, these large $q$ settings are actually those where state-of-the-art \textsc{mcmc} methods, including \textsc{stan} implementations of Hamiltonian no-u-turn samplers \citep{hoff_2014}, are computationally  unfeasible. The results in \citet{botev_2017} are also useful to compute efficiently Gaussian cumulative distribution functions, and hence are practically relevant to evaluate  \eqref{eq10} and   \eqref{eq11} in small-to-moderate $h+m$ settings.

\subsubsection{Blocked Partially-Factorized Variational Bayes}
\label{sec.3.2.2}
As discussed in Section \ref{sec.3.2.1}, when $h+m$ is large, sampling from $(h+m)$-variate truncated normals with full covariance matrix becomes computationally unfeasible  \citep{botev_2017}, thus making Algorithm \ref{algo:samplingIID} impractical in these settings. Typically, $h$ is either $0$---when Gaussian priors are considered---or is equal to a small value, whereas $m$ depends on the sample size $n$ and on the number of classes $L$; see Table~\ref{t1}. Hence, it is necessary to devise more scalable methods, especially in common settings where $n$  is larger than a few hundreds.

A possible solution to the above problem is to consider approximations of the posterior density, with variational Bayes providing a well-established procedure, especially in those models admitting simple augmented data representations \citep{blei2017}. As clarified in Section \ref{sec.2},  this is the case of multinomial probit models relying on  Gaussian latent utilities. Such a property has motivated several variational strategies to approximate the joint posterior $p(\bbeta, \bar{\bz} \mid \by, \bX)$ of  $\bbeta$  and the augmented data $\bar{\bz}$,  with a  tractable density $q^*(\bbeta, \bar{\bz})$, which  is the closest in Kullback--Leibler (\textsc{kl}) divergence \citep{kullback_1951} to $p(\bbeta, \bar{\bz} \mid \by, \bX)$, among all the densities which belong to a pre-specified approximating family $\mathcal{Q}$. As for the development of simple Gibbs samplers relying on tractable full-conditionals \citep{Albert_1993}, the inclusion of the augmented data facilitates the implementation of simple coordinate ascent variational inference (\textsc{cavi}) routines \citep{bishop2006,blei2017} to minimize, with respect to $q(\bbeta, \bar{\bz})$, the divergence $\textsc{kl}[q(\bbeta, \bar{\bz})||p(\bbeta, \bar{\bz} \mid \by, \bX)]$. 

Clearly, the availability of simple optimization routines and strategies to derive the optimal marginal $q^*(\bbeta)$ from $q^*(\bbeta, \bar{\bz})$, depend also the choice of the family $\mathcal{Q}$. Common solutions in binary \citep[e.g.,][]{consonni_2007} and multinomial \citep[e.g.,][]{girolami2006}  probit settings rely on mean-field families $\mathcal{Q}_{\textsc{mf}}=\{ q(\bbeta, \bar{\bz}): q(\bbeta, \bar{\bz})=q(\bbeta)q(\bar{\bz})\}$ that assume independence between $\bbeta$ and $\bar{\bz}$. These strategies come with simple \textsc{cavi} algorithms which scale easily to high-dimensional settings and, due to the factorized form of $q^*(\bbeta, \bar{\bz})$, provide as a byproduct the approximating density $q^*(\bbeta)$ of direct interest. However, recent theoretical and empirical studies on simple univariate  probit models \citep{fasano2019asymptotically}, have shown that such a mean-field assumption often leads to a low-quality approximation in high-dimensional probit settings, which severely affects not only uncertainty quantification, but also estimation and classification. To address this issue in the context of basic univariate probit regression with Gaussian priors, \citet{fasano2019asymptotically} considered a partially-factorized mean-field approximating family $\mathcal{Q}_{\textsc{pfm}}=\{ q(\bbeta, \bar{\bz}): q(\bbeta, \bar{\bz})=q(\bbeta \mid\bar{\bz} ) \prod_{r=1}^{h+m}q(\bar{z}_r)\}$ which avoids enforcing independence between $\bbeta$ and $\bar{\bz}$, and only assumes that $q(\bar{\bz})$ factorizes as the product of its marginals. This novel class of approximating densities substantially improves the quality of the original mean-field approximation and almost perfectly matches the exact posterior in high-dimensional settings, especially when the number of predictors is higher than the sample size, without sacrificing computational tractability. Unfortunately, this strategy is only available for univariate binary probit models with Gaussian priors.

Motivated by the above discussion, we develop a new blocked partially-factorized mean-field approximation which extends the contribution of \citet{fasano2019asymptotically} in three main important directions. In particular, we [i] allow the inclusion of \textsc{sun} and not only Gaussian priors, [ii] generalize the methods to multinomial probit models, and [iii] further enlarge the class of approximating densities by replacing $ \prod_{r=1}^{h+m}q(\bar{z}_r)$ in $\mathcal{Q}_{\textsc{pfm}}$ with $ \prod_{c=1}^{C} q(\bar{\bz}_c)$, where $\bar{\bz}_1, \ldots, \bar{\bz}_{C}$ are distinct sub-vectors of $\bar{\bz}$, such that $\bar{\bz}=(\bar{\bz}^{\intercal}_1, \ldots, \bar{\bz}^{\intercal}_{C})^{\intercal}$. Therefore, instead of enforcing independence among all the augmented data, we only make this assumption between pre-specified blocks.  In fact, while in high-dimensional univariate binary settings the independence among all the augmented data does not seem to have a major impact on the quality of the approximation  \citep{fasano2019asymptotically}, this may not be the case in multinomial probit models. For example, under the formulation presented  in Section \ref{sec.2.2}, every  unit $i$ enters the matrix $\bar{\bX}$ multiple times and, hence, it is reasonable to expect a relatively strong dependence among unit-specific augmented data, which cannot be accurately approximated by a fully factorized representation for $q(\bar{\bz})$. Similar blocking  ideas have been also considered by \citet{chop_2011,genton2018} and \citet{cao2019}, to simulate from multivariate truncated normals and compute cumulative distribution functions of high-dimensional Gaussians. We adapt these ideas in the context of variational inference to obtain improved approximations of the posterior, without affecting computational performance. 

To introduce the blocked partially-factorized  mean-field approximation, first note that the kernel of the posterior density $p(\bbeta \mid \by, \bX)$ in \eqref{eq9} can be re-written as
\begin{eqnarray}
p(\bbeta \mid \by, \bX) \propto \phi_q(\bbeta-\bxi_{\post};\bOmega_{\post})\int \phi_{h{+}m}(\bar{\bz}-(\boeta_{\post}+\bX_{\post}\bbeta); \bSigma_{\post}) \mathbbm{1}(\bar{\bz}>\bzero) \mbox{d} \bar{\bz},
\label{eq12}
\end{eqnarray}
where $\bX_{\post}=\bDelta_{\post}^{\intercal}\bar{\bOmega}_{\post}^{-1}\bomega_{\post}^{-1}$, $\boeta_{\post}=\bgamma_{\post}-\bX_{\post}\bxi_{\post}$,  and $\bSigma_{\post}=\bGamma_{\post}-\bDelta_{\post}^{\intercal}\bar{\bOmega}_{\post}^{-1}\bDelta_{\post}$. To clarify the connection between  \eqref{eq9} and  \eqref{eq12} it suffices to note that the integral in \eqref{eq12}  actually coincides with the multivariate Gaussian cumulative distribution function $\Phi_{h+m}(\boeta_{\post}+\bX_{\post}\bbeta; \bSigma_{\post})$ in  the numerator of \eqref{eq9}. Leveraging this alternative representation and Gaussian--Gaussian conjugacy, we can easily notice that
\begin{eqnarray}
\begin{split}
p(\bbeta \mid \bar{\bz},\by, \bX) &\propto \phi_q(\bbeta-\bxi_{\post};\bOmega_{\post})\phi_{h{+}m}(\bar{\bz}-(\boeta_{\post}+\bX_{\post}\bbeta); \bSigma_{\post}),\\
 & \propto  \phi_q(\bbeta-\bV_{\post}[\bX^{\intercal}_{\post}\bSigma_{\post}^{-1}(\bar{\bz}-\boeta_{\post})+\bOmega^{-1}_{\post}\bxi_{\post}]; \bV_{\post}),
 \end{split}
\label{eq13}
\end{eqnarray}
where $\bV_{\post}=(\bX^{\intercal}_{\post}\bSigma^{-1}_{\post}\bX_{\post}+\bOmega^{-1}_{\post})^{-1}$. Hence, $(\bbeta \mid \bar{\bz},\by, \bX) \sim \mbox{N}_q(\bV_{\post}[\bX^{\intercal}_{\post}\bSigma_{\post}^{-1}(\bar{\bz}-\boeta_{\post})+\bOmega^{-1}_{\post}\bxi_{\post}], \bV_{\post})$. On the other hand, according to \eqref{eq12}, the conditional density $p(\bar{\bz} \mid\bbeta, \by, \bX)$ of the augmented data $\bar{\bz}$ is a multivariate truncated normal with mean $\boeta_{\post}+\bX_{\post}\bbeta$, covariance matrix $\bSigma_{\post}$ and truncation below $\bzero$. Therefore, marginalizing out $\bbeta$ with density $\phi_q(\bbeta-\bxi_{\post};\bOmega_{\post})$, yields
\begin{eqnarray}
\begin{split}
p(\bar{\bz} \mid\by, \bX) &\propto \phi_{h{+}m}(\bar{\bz}-(\boeta_{\post}+\bX_{\post}\bxi_{\post}); \bSigma_{\post}+\bX_{\post}\bOmega_{\post}\bX^{\intercal}_{\post}) \mathbbm{1}(\bar{\bz}>\bzero),\\
&\propto \phi_{h{+}m}(\bar{\bz}-\bgamma_{\post}; \bGamma_{\post})\mathbbm{1}(\bar{\bz}>\bzero),
\end{split}
\label{eq14}
\end{eqnarray}
since $\boeta_{\post}=\bgamma_{\post}{-} \bX_{\post}\bxi_{\post}$ and $\bGamma_{\post}=\bSigma_{\post}{+}\bX_{\post}\bOmega_{\post}\bX^{\intercal}_{\post}$. Combining \eqref{eq13}--\eqref{eq14} and recalling previous discussion, we aim to obtain an accurate approximation $q^*(\bbeta,\bar{\bz} )$ of the joint density
\begin{eqnarray}
\begin{split}
&p(\bbeta,\bar{\bz} \mid\by, \bX)=p(\bbeta \mid \bar{\bz},\by, \bX) p(\bar{\bz} \mid\by, \bX),\\
&\propto \phi_q(\bbeta-\bV_{\post}[\bX^{\intercal}_{\post}\bSigma_{\post}^{-1}(\bar{\bz}-\boeta_{\post})+\bOmega^{-1}_{\post}\bxi_{\post}]; \bV_{\post}) \phi_{h{+}m}(\bar{\bz}-\bgamma_{\post}; \bGamma_{\post}) \mathbbm{1}(\bar{\bz}>\bzero), 
 \end{split}
\label{eq15}
\end{eqnarray}
such that $q^*(\bbeta,\bar{\bz} )$ minimizes the \textsc{kl} divergence $\textsc{kl}[q(\bbeta, \bar{\bz})||p(\bbeta, \bar{\bz} \mid \by, \bX)]$ within the blocked partially-factorized mean-field family $\mathcal{Q}_{\textsc{pfm-b}}=\{ q(\bbeta, \bar{\bz}): q(\bbeta, \bar{\bz})=q(\bbeta \mid\bar{\bz} ) \prod_{c=1}^{C} q(\bar{\bz}_c)\}$, where $\bar{\bz}_1, \ldots, \bar{\bz}_C$ are the pre-specified sub-vectors of $\bar{\bz}$. Formulation \eqref{eq15} clarifies why $\mathcal{Q}_{\textsc{pfm-b}}$ provides a particularly suitable family of approximating densities for $p(\bbeta,\bar{\bz} \mid\by, \bX)$. In particular, since the exact conditional density $p(\bbeta \mid \bar{\bz},\by, \bX)$ has a tractable Gaussian form, assuming independence between $\bbeta$ and $\bar{\bz}$ as in classical mean-field variational Bayes seems an unnecessarily strong assumption. On the other hand, the main source of intractability in $p(\bbeta,\bar{\bz} \mid\by, \bX)$ arises from the high-dimensional truncated normal density $p(\bar{\bz} \mid\by, \bX)$ with full covariance matrix $\bGamma_{\post}$, thus motivating our attempt to approximate it via a set of $C$ independent lower-dimensional truncated normal densities $q^*(\bar{\bz}_1)\cdots q^*(\bar{\bz}_C)$. Each of these blocks must be sufficiently small to allow tractable inference under the associated truncated normal  approximation, and should be specified so as to group augmented data with strong correlations in $\bGamma_{\post}$. Remark \ref{remark1} discusses and motivates a possible default strategy to define the different blocks in multinomial probit models, when necessary.

\begin{Remark}
In multinomial probit models, when necessary, it is typically sufficient to  group augmented data associated with the same unit $i$, provided that there may be strong overlap in the rows of $\bar{\bX}$ referring to $i$, thereby leading to high correlation in $\bGamma_{\post}$. This choice is further motivated by the fact that the optimal mean-field solution---which does not assume factorized forms for $q(\bar{\bz})$ in $\mathcal{Q}_{\textsc{mf}}$---is defined as $q_{\textsc{mf}}^*(\bbeta,\bar{\bz})=q^*_{\textsc{mf}}(\bbeta) \prod_{i=1}^n q^*_{\textsc{mf}}(\bar{\bz}_i)$ under Gaussian priors \citep{girolami2006}. Such a solution belongs also to $\mathcal{Q}_{\textsc{pfm-b}}$ when blocking according to $i$. Therefore, $\mbox{min}_{q(\bbeta,\bar{\bz}) \in \mathcal{Q}_{\textsc{pfm-b}}}\textsc{kl}[q(\bbeta, \bar{\bz})||p(\bbeta, \bar{\bz} \mid \by, \bX)]\leq \mbox{min}_{q(\bbeta,\bar{\bz}) \in \mathcal{Q}_{\textsc{mf}}}\textsc{kl}[q(\bbeta, \bar{\bz})||p(\bbeta, \bar{\bz} \mid \by, \bX)]$. Moreover, since $\mathcal{Q}_{\textsc{pfm}} \subset \mathcal{Q}_{\textsc{pfm-b}}$, we also have that $\mbox{min}_{q(\bbeta,\bar{\bz}) \in \mathcal{Q}_{\textsc{pfm-b}}}\textsc{kl}[q(\bbeta, \bar{\bz})||p(\bbeta, \bar{\bz} \mid \by, \bX)]\leq \mbox{min}_{q(\bbeta,\bar{\bz}) \in \mathcal{Q}_{\textsc{pfm}}}\textsc{kl}[q(\bbeta, \bar{\bz})||p(\bbeta, \bar{\bz} \mid \by, \bX)]$. Hence, when blocking according to $i$, our solution is guaranteed to improve mean-field variational Bayes and recent partially-factorized extensions, under Gaussian priors. Similar arguments can be made under \textsc{sun} priors.
\label{remark1}
\end{Remark}

Besides providing a wider and more flexible class, the family $\mathcal{Q}_{\textsc{pfm-b}}$ also allows straightforward optimization, as shown in Proposition \ref{prop4}.

\begin{Proposition}
The \textsc{kl} divergence $\textsc{kl}[q(\bbeta, \bar{\bz})||p(\bbeta, \bar{\bz} \mid \by, \bX) ]$ between $q(\bbeta, \bar{\bz}) \in \mathcal{Q}_{\textsc{pfm-b}}$ and $p(\bbeta, \bar{\bz} \mid \by, \bX)$ in \eqref{eq15}, is minimized at $q^*(\bbeta, \bar{\bz})=q^*(\bbeta \mid\bar{\bz} ) \prod_{c=1}^{C} q^*(\bar{\bz}_c)$, with
\begin{eqnarray}
q^*(\bbeta \mid\bar{\bz} )&\propto&  \phi_q(\bbeta-\bV_{\post}[\bX^{\intercal}_{\post}\bSigma_{\post}^{-1}(\bar{\bz}-\boeta_{\post})+\bOmega^{-1}_{\post}\bxi_{\post}]; \bV_{\post}),
\label{eq16}\\
 q^*(\bar{\bz}_c) &\propto& \phi_{n_c}(\bar{\bz}_c-\bgamma_{\post[c]}-\bW_{\post[c]}(\mathbb{E}_{q^*(\bar{\bz}_{-c})}(\bar{\bz}_{-c})-\bgamma_{\post[-c]});\bGamma_{\post[c]})\mathbbm{1}(\bar{\bz}_c>\bzero), \ \forall c,  \qquad
 \label{eq17}
\end{eqnarray}
where $\bW_{\post[c]}=\bGamma_{\post[c,-c]}(\bGamma_{\post[-c,-c]})^{-1}$ and $\bGamma_{\post[c]}=\bGamma_{\post[c,c]}-\bGamma_{\post[c,-c]}(\bGamma_{\post[-c,-c]})^{-1}\bGamma_{\post[-c,c]}$, with $\bGamma_{\post[c,c]}$, $\bGamma_{\post[-c,-c]}$, $\bGamma_{\post[-c,c]}$ and $\bGamma_{\post[c,-c]}$ denoting the four blocks of  $\bGamma_{\post}$ when partitioned to highlight  sub-vector $\bar{\bz}_c$ against  all the others in $\bar{\bz}_{-c}$. Similarly, $\bgamma_{\post[c]}$ and $\bgamma_{\post[-c]}$ are the sub-vectors of $\bgamma_{\post}$ referring to block $c$ and the remaining blocks, respectively. Finally, 
$\mathbb{E}_{q^*(\bar{\bz}_{-c})}(\bar{\bz}_{-c}){=}(\mathbb{E}_{q^*(\bar{\bz}_{1})}(\bar{\bz}_{1})^{\intercal}, \ldots,\mathbb{E}_{q^*(\bar{\bz}_{c-1})}(\bar{\bz}_{c-1})^{\intercal},\mathbb{E}_{q^*(\bar{\bz}_{c+1})}(\bar{\bz}_{c+1})^{\intercal}, \ldots, \mathbb{E}_{q^*(\bar{\bz}_{C})}(\bar{\bz}_{C})^{\intercal})^{\intercal}$, where the expectations are taken with respect to the optimal truncated normal approximations.
\label{prop4}
\end{Proposition}

\begin{algorithm*}[t]
	\caption{\textsc{cavi} for blocked partially-factorized  approximation in Proposition~\ref{prop4}} 
		\For{$t = 1$ until convergence of the \textsc{elbo}}{
\For{$c = 1, \ldots, C$}{
Set   $\mathbb{E}_{q^{(t)}(\bar{\bz}_{c})}(\bar{\bz}_{c})$ equal to the expected value of an $n_c$-variate Gaussian with mean $\bgamma_{\post[c]}+\bW_{\post[c]}(\mathbb{E}_{q^{(t-1)}(\bar{\bz}_{-c})}(\bar{\bz}_{-c})-\bgamma_{\post[-c]})$, covariance matrix $\bGamma_{\post[c]}$ and truncation below $\bzero$, where  $\mathbb{E}_{q^{(t-1)}(\bar{\bz}_{-c})}(\bar{\bz}_{-c})$ is defined as $$(\mathbb{E}_{q^{(t)}(\bar{\bz}_{1})}(\bar{\bz}_{1})^{\intercal}, \ldots,\mathbb{E}_{q^{(t)}(\bar{\bz}_{c-1})}(\bar{\bz}_{c-1})^{\intercal},\mathbb{E}_{q^{(t-1)}(\bar{\bz}_{c+1})}(\bar{\bz}_{c+1})^{\intercal}, \ldots, \mathbb{E}_{q^{(t-1)}(\bar{\bz}_{C})}(\bar{\bz}_{C})^{\intercal})^{\intercal}.$$ [in \texttt{R} use function \texttt{MomTrunc} to compute the mean of truncated normals].
	}}
	{\bf Output:} Optimal truncated normal approximating densities $q^*(\bar{\bz}_1), \ldots, q^*(\bar{\bz}_C)$ from \eqref{eq17}, which are  combined with the closed-form solution for $q^*(\bbeta \mid\bar{\bz})$ in~\eqref{eq16}, to provide the optimal joint approximate density  $q^*(\bbeta, \bar{\bz})=q^*(\bbeta \mid\bar{\bz} ) \prod_{c=1}^{C} q^*(\bar{\bz}_c)$.	
		\label{algo:2}
\end{algorithm*}

The solution in  \eqref{eq16} is a direct consequence of the chain rule for the \textsc{kl} divergence. In fact, $\textsc{kl}[q(\bbeta, \bar{\bz})||p(\bbeta, \bar{\bz} \mid \by, \bX) ]=\textsc{kl}[q(\bar{\bz})||p(\bar{\bz} \mid \by, \bX) ]+\mathbb{E}_{q(\bar{\bz})}\{\textsc{kl}[q(\bbeta \mid\bar{\bz} )||p(\bbeta \mid \bar{\bz},\by, \bX)] \}$, and hence the non-negative second summand is exactly zero for every $q(\bar{\bz})$ only when $q^*(\bbeta \mid\bar{\bz})=p(\bbeta \mid \bar{\bz},\by, \bX)$. To clarify the result in \eqref{eq17}, recall that the optimal solution for $q(\bar{\bz}_c)$ is proportional to $\exp[\mathbb{E}_{q^*(\bar{\bz}_{-c})}(\log[p(\bar{\bz}_c \mid \bar{\bz}_{-c},\by,\bX)])]$ \citep{bishop2006,blei2017}. Hence, since $(\bar{\bz} \mid\by, \bX)$ has a multivariate truncated Gaussian density in \eqref{eq14}, it follows that also each $p(\bar{\bz}_c \mid \bar{\bz}_{-c},\by,\bX)$ is an $n_c$-variate truncated normal density, whose log-kernel is linear in $\bar{\bz}_{-c}$ and the remaining parameters are specified as in \eqref{eq17} \citep[e.g.,][]{horrace2005,holmes_2006}. According to Proposition \ref{prop4}, the only unknown parameters are $\mathbb{E}_{q^*(\bar{\bz}_{c})}(\bar{\bz}_{c})$, $c=1, \ldots, C$, whose solution requires solving a non-linear system of equations. Algorithm~\ref{algo:2} summarizes the steps of the \textsc{cavi} to obtain such quantities via simple operations.
 
Once $q^*(\bbeta \mid\bar{\bz} )$ and  $q^*(\bar{\bz})=\prod_{c=1}^{C} q^*(\bar{\bz}_c)$ are available, approximations of key functionals of $\bbeta$ can be easily derived leveraging the law of total expectation and results in Proposition \ref{prop4}. In particular, since $\mathbb{E}_{q^*(\bbeta)}(\bbeta)=\mathbb{E}_{q^*(\bar{\bz})}[\mathbb{E}_{q^*(\bbeta \mid\bar{\bz} )}(\bbeta)]$, we have that
\begin{eqnarray}
\mathbb{E}_{q^*(\bbeta)}(\bbeta)=\bV_{\post}[\bX^{\intercal}_{\post}\bSigma_{\post}^{-1}(\mathbb{E}_{q^*(\bar{\bz})}(\bar{\bz})-\boeta_{\post})+\bOmega^{-1}_{\post}\bxi_{\post}],
\label{eq18}
\end{eqnarray}
whereas, the equality $\mbox{var}_{q^*(\bbeta)}(\bbeta)=\mathbb{E}_{q^*(\bar{\bz})}[\mbox{var}_{q^*(\bbeta \mid\bar{\bz} )}(\bbeta)]+\mbox{var}_{q^*(\bar{\bz})}[\mathbb{E}_{q^*(\bbeta \mid\bar{\bz} )}(\bbeta)]$, leads to
\begin{eqnarray}
\mbox{var}_{q^*(\bbeta)}(\bbeta)=\bV_{\post}+\bV_{\post}\bX^{\intercal}_{\post}\bSigma_{\post}^{-1}\mbox{var}_{q^*(\bar{\bz})}(\bar{\bz})\bSigma_{\post}^{-1}\bX_{\post}\bV_{\post}.
\label{eq19}
\end{eqnarray}
To evaluate \eqref{eq18} and \eqref{eq19}, it is sufficient to compute $\mathbb{E}_{q^*(\bar{\bz}_c )}(\bar{\bz}_c)$ and $\mbox{var}_{q^*(\bar{\bz}_c )}(\bar{\bz}_c)$, separately for each $c=1, \ldots, C$, since, due to the independence assumption among the $C$ sub-vectors of $\bar{\bz}$, the vector $\mathbb{E}_{q^*(\bar{\bz})}(\bar{\bz})$ has blocks $\mathbb{E}_{q^*(\bar{\bz})}(\bar{\bz})_{[c]}=\mathbb{E}_{q^*(\bar{\bz}_c )}(\bar{\bz}_c)$, for each $c=1, \ldots, C$, whereas $\mbox{var}_{q^*(\bar{\bz})}(\bar{\bz})$ is a block diagonal matrix with blocks $\mbox{var}_{q^*(\bar{\bz})}(\bar{\bz})_{[cc]}=\mbox{var}_{q^*(\bar{\bz}_c )}(\bar{\bz}_c)$. Recalling Remark~\ref{remark1}, in multinomial probit models such blocks often refer to rows in the design matrix $\bar{\bX}$ corresponding to the same unit $i$ and, hence, their dimensions are, by definition, equal or lower than the number of classes $L$, which is small in most applications. This allows rapid  evaluation of $\mathbb{E}_{q^*(\bar{\bz}_c )}(\bar{\bz}_c)$ and $\mbox{var}_{q^*(\bar{\bz}_c )}(\bar{\bz}_c)$ via routine \texttt{R} functions such as \texttt{MomTrunc}.

Although \eqref{eq18} and \eqref{eq19} are typically the main quantities of interest, other generic functionals $\mathbb{E}_{q^*(\bbeta)}[g(\bbeta)]$ can be easily derived via simple Monte Carlo methods based on samples from  $q^*(\bbeta)$. Combining \eqref{eq16}--\eqref{eq17}, such draws can be obtained by setting
\begin{eqnarray}
\bbeta^{(t)}=\bV_{\post}[\bX^{\intercal}_{\post}\bSigma_{\post}^{-1}([\bar{\bz}^{(t) \intercal}_1, \ldots, \bar{\bz}^{(t) \intercal}_C]^{\intercal}-\boeta_{\post})+\bOmega^{-1}_{\post}\bxi_{\post}]+\beps^{(t)}, \quad t=1, \ldots, T,
\label{eq20}
\end{eqnarray}
where $\beps^{(t)} \sim \mbox{N}_q(\bzero, \bV_{\post})$, and $\bar{\bz}^{(t)}_c \sim \mbox{TN}_{n_c}(\bzero; \bgamma_{\post[c]}+\bW_{\post[c]}(\mathbb{E}_{q^*(\bar{\bz}_{-c})}(\bar{\bz}_{-c})-\bgamma_{\post[-c]}), \bGamma_{\post[c]})$ for $c=1, \ldots, C$. Also in this case, since $n_c$ is typically very small, samples from $n_c$-variate truncated normals can be effectively obtained from common \texttt{R} functions, such as \texttt{rtmvnorm}. This  strategy is particularly useful to compute the predictive probabilities for a new unit with covariates $\bx_{\new}$. To accomplish this goal, it suffices to compute, for each sample $\bbeta^{(t)}$ of $\bbeta$, the latent utilities $z^{_{(t)}}_{\new l}$, $l=1, \ldots, L$ defined either via \eqref{eq1}, \eqref{eq3} or \eqref{eq5}, depending on the multinomial probit model considered. Then, if the focus is on models \eqref{eq1} and \eqref{eq3}, a Monte Carlo estimate for $\pr(y_{\new}=l \mid \by, \bX, \bx_{\new})$ can be obtained by computing the relative frequency of samples in which $z^{_{(t)}}_{\new l}>z^{_{(t)}}_{\new k}$ for all $k \neq l$. If, instead, one considers the sequential representation in  \eqref{eq5},   the Monte Carlo estimate for $\pr(y_{\new}=l \mid \by, \bX, \bx_{\new})$ coincides with the relative frequency of samples in which $z^{_{(t)}}_{\new l}>0$ and $z^{_{(t)}}_{\new k}<0$, for $k<l$.

\section{Simulation Study}
To evaluate the performance of the computational methods presented in Section \ref{sec.3.2} relative to state-of-the-art competitors, we consider an extensive simulation study under different settings of $p$ and $L$. The main objective is to assess empirical evidence of improved accuracy and computational scalability for Algorithms~\ref{algo:samplingIID} and \ref{algo:2}, over routinely-implemented mean-field approximations \citep[e.g.,][]{girolami2006, consonni_2007}, and Hamiltonian Monte Carlo sampling schemes \citep[e.g.,][]{hoff_2014} under the \texttt{rstan} library. The latter \textsc{mcmc} strategy provides an accurate but expensive default solution in high dimensions, and, hence, is taken as a benchmark throughout the simulation study.

As discussed in Section~\ref{sec.3.2}, the gains provided by the proposed computational methods are valid for all the three multinomial probit models presented in Sections~\ref{sec.2.1}--\ref{sec.2.3}. Due to this and consistent with the application in Section~\ref{sec.5}, we consider the sequential multinomial probit model in Section \ref{sec.2.3} as a representative example to quantify empirically the magnitude of these gains at varying combinations of $p\in\{75; 100; 200\}$ and $L\in \{3; 5; 7; 9\}$. The sample size is, instead, kept fixed and equal to $n=100$ consistent with the empirical results in \cite{fasano2019asymptotically} for basic univariate probit regression  which show that the accuracy of variational strategies mainly depends on $p$ and $n$ through the ratio among such quantities. It is also worth noticing that the settings for $n$, $p$ and $L$ considered in this simulation are much lower than those which could be handled  under the blocked partially-factorized and mean-field approximations. Such moderate dimensions of $n$, $p$ and $L$ are required to avoid possible computational bottlenecks in obtaining i.i.d.\ samples from Algorithm \ref{algo:samplingIID} and \textsc{mcmc} draws under \texttt{rstan}, respectively.

\label{sec.4}
\begin{figure}[t]
	\centering
	\includegraphics[width=15.7cm, height=0.52\textheight]{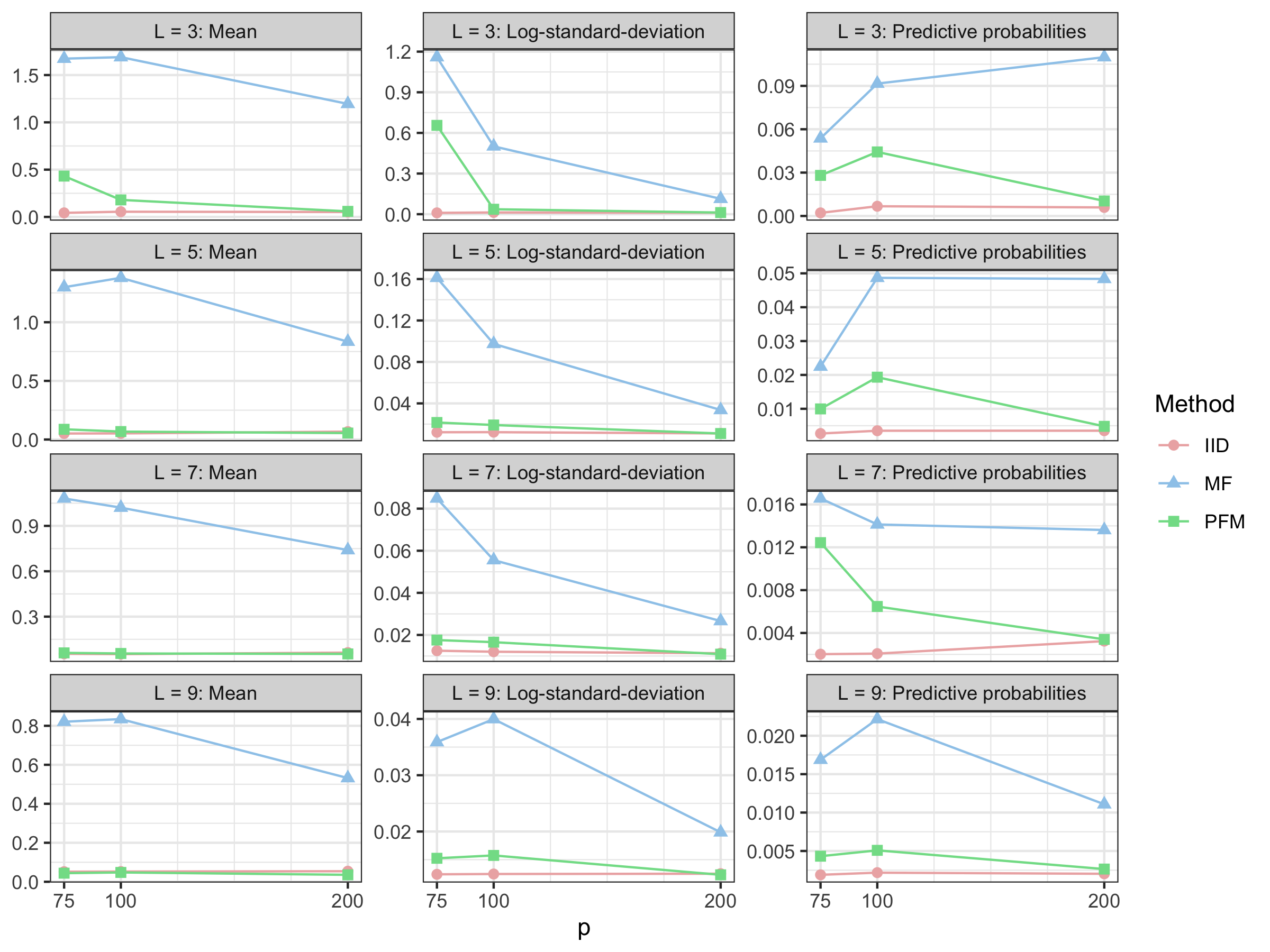}
	\caption{For  three functionals of main interest and different settings of $L \in \{3;5;7;9\}$, trajectories for the median of the absolute differences, at varying $p  \in \{75;100;200\}$, between an accurate but expensive Monte Carlo evaluation---via \texttt{rstan}---of such functionals and their estimates provided by the i.i.d.\ sampling scheme from the exact posterior presented in Section~\ref{sec.3.2.1}  (\textsc{iid}), the blocked partially-factorized variational Bayes proposed in Section~\ref{sec.3.2.2} (\textsc{pfm}), and classical mean-field variational Bayes (\textsc{mf}), respectively.  }
	\label{fig:F0}
\end{figure}

For each combination $(p,L)$, the predictors in $\bX$ are simulated from independent $\mbox{N}(0,1)$ variables and, as suggested in \citet{gelman2008} and \citet{chopin_2017}, such predictors are subsequently standardized to have mean zero and standard deviation $0.5$, for the training data. The coefficients $\beta_{lj}$, $l=1,\ldots,L-1$, $j=1,\ldots,p$ comprising the vector $\bbeta$ are, instead, generated  independently  from a uniform distribution in $(-5,5)$. Leveraging the realizations of $\bX$ and $\bbeta$, the categorical responses in $\by$ are simulated from the sequential Bernoulli choice mechanism outlined in Section~\ref{sec.2.3}. To assess the quality in classification, we also simulate 20 test units following the same procedure presented for the $n=100$ training data. Consistent with common implementations of regression models for binary or categorical responses \citep[e.g.,][]{gelman2008, chopin_2017}, Bayesian inference  is performed under independent weakly informative Gaussian priors for the coefficients in $\bbeta$, with zero mean and variance $\omega^2 = 25$. 

To evaluate the performance of the proposed methods, we conduct posterior inference under the strategies developed in Sections \ref{sec.3.2.1}--\ref{sec.3.2.2} and compare the results against state-of-the-art alternatives comprising classical mean-field approximations \citep[e.g.,][]{consonni_2007,girolami2006}, and the \texttt{rstan} implementation of Hamiltonian Monte Carlo \citep[e.g.,][]{hoff_2014}. More specifically, we consider as benchmark posterior inference for selected functionals of interest computed from  $5000$ \texttt{rstan} samples, and compare such quantities with those resulting from $5000$ i.i.d.\ samples from the exact \textsc{sun} posterior under Algorithm \ref{algo:samplingIID}, and the ones computed from the approximate densities provided by the blocked  partially-factorized  strategy  in Algorithm \ref{algo:2}, and the classical mean-field variational Bayes solution. Figure \ref{fig:F0} summarizes the output of this comparison, with a focus on posterior means, standard deviations, and predictive probabilities for the 20 test units. For these quantities, we display the median of the absolute differences between the \texttt{rstan} estimates and those arising from the other three strategies under analysis. In the first two panels, the medians are computed from the $p(L-1)$ absolute differences for the estimates of the posterior moments for every $\beta_{lj}$, $l=1,\ldots,L-1$, $j=1,\ldots,p$, obtained under the different methods, while in the third panel such quantities are calculated for the $20 L$ predictive probabilities, estimated for each test unit and category.

As shown in Figure \ref{fig:F0}, the  partially-factorized solution yields uniformly improved accuracy relative to the classical mean-field one, and the quality of the approximation increases with both $p$ and $L$. Moreover, the error rapidly vanishes when the dimension of $\bbeta$ exceeds $n$. These results are coherent with the empirical findings in \citet{fasano2019asymptotically} on classical univariate binary probit models. Such a superior performance comes at almost no expenses in computational budget, since the average runtime required to obtain the functionals of interest under blocked partially-factorized and mean-field approximations is, respectively, $2.9$  and $2.5$ seconds, with a maximum of $11.3$ and $10.6$ seconds, respectively, under the scenario $(p,L)=(200,9)$. These runtimes are orders of magnitude faster than those of the \texttt{rstan} implementation of Hamiltonian Monte Carlo, which, on average, requires $489$ seconds. Indeed, although \texttt{rstan} is efficient in low dimensions, such a method faces increasing computational difficulties as $p$ and $L$ grow, thereby yielding average runtimes of approximately $900$ seconds in $p>n$ scenarios. These are exactly the settings in which the i.i.d\ sampler described in Algorithm \ref{algo:samplingIID}---that provides similarly accurate estimates relative to \texttt{rstan}---displays the highest computational advantages, with  an average runtime of only $33$ seconds. Therefore, Algorithm \ref{algo:samplingIID} addresses a gap in the literature regarding posterior inference in multinomial probit models with small-to-moderate sample size and high-dimensional coefficients vector, a setting where state-of-the-art \textsc{mcmc} are computationally inefficient. When $n$ grows, the blocked partially-factorized approximation described in Section \ref{sec.3.2.2} and implemented in Algorithm \ref{algo:2} provides a practically feasible and effective solution which uniformly improves the accuracy of standard mean-field strategies, and yields almost the same estimates of state-of-the-art sampling methods when $p(L-1)>n$, at massively lower runtimes. These results on accuracy remained consistent  also when comparing other quantiles of the absolute differences. 

\section{Gastrointestinal Lesions Application}
\label{sec.5}
To confirm  findings in Section \ref{sec.4} also on a real-world application, we consider a medical study by \citet{mesejo2016} that focuses on  $76$ gastrointestinal lesions classified as \texttt{hyperplastic} ($l=1$),  \texttt{serrated adenoma} ($l=2$) and  \texttt{adenoma} $(l=3)$ where the first is  benign, whereas the others are malignant. For every individual lesion, a vector of $1396$ features is available, and comprises \textsc{2d} textural, \textsc{2d} color, and \textsc{3d} shape measurements, collected with white light and narrow band imaging. In our study we first remove the features that are always $0$, and then standardize the remaining ones as suggested by \citet{gelman2008} and \citet{chopin_2017}, thus obtaining $p-1=929$ predictors  with mean $0$ and standard deviation $0.5$.  To assess predictive performance, we also hold out $15$ randomly chosen units from the calculation of the posterior, roughly corresponding to $20\%$ of the observations.

As discussed in Section \ref{sec.1}, Bayesian inference for such a high-dimensional study is computationally unfeasible under state-of-the-art \textsc{mcmc} methods \citep{chopin_2017}, and hence it provides a useful setting for quantifying to what extent our  results  in Section~\ref{sec.3} can cover this gap. To address such a goal, we first focus on the sequential discrete choice multinomial probit model in Section \ref{sec.2.3} with Gaussian priors, and compare  the computational performance of the methods developed in Section \ref{sec.3.2} with the \texttt{rstan} implementation of the Hamiltonian no-u-turn sampler in \citet{hoff_2014}.  The choice of the sequential model is directly motivated by the type of response of interest in our study. Indeed, it is plausible to first model benign ($l=1$) against malignant ($l>1$) status, and then focus on comparing the two sub-categories $l=2$ and $l=3$ of malignant lesions. Under this model, the vector $\bbeta$ has dimension $1860$, corresponding to the two class-specific $929$-dimensional parameter vectors plus a class-specific intercept term. Consistent with the simulation study in Section \ref{sec.4}, we place a $\mbox{N}_{1860}(\boldsymbol{0},\omega^2\bI_{1860})$ prior on $\bbeta$, with $\omega^2=25$  \citep{gelman2008}.

\begin{figure}[t]
	\centering
		\includegraphics[scale=.198]{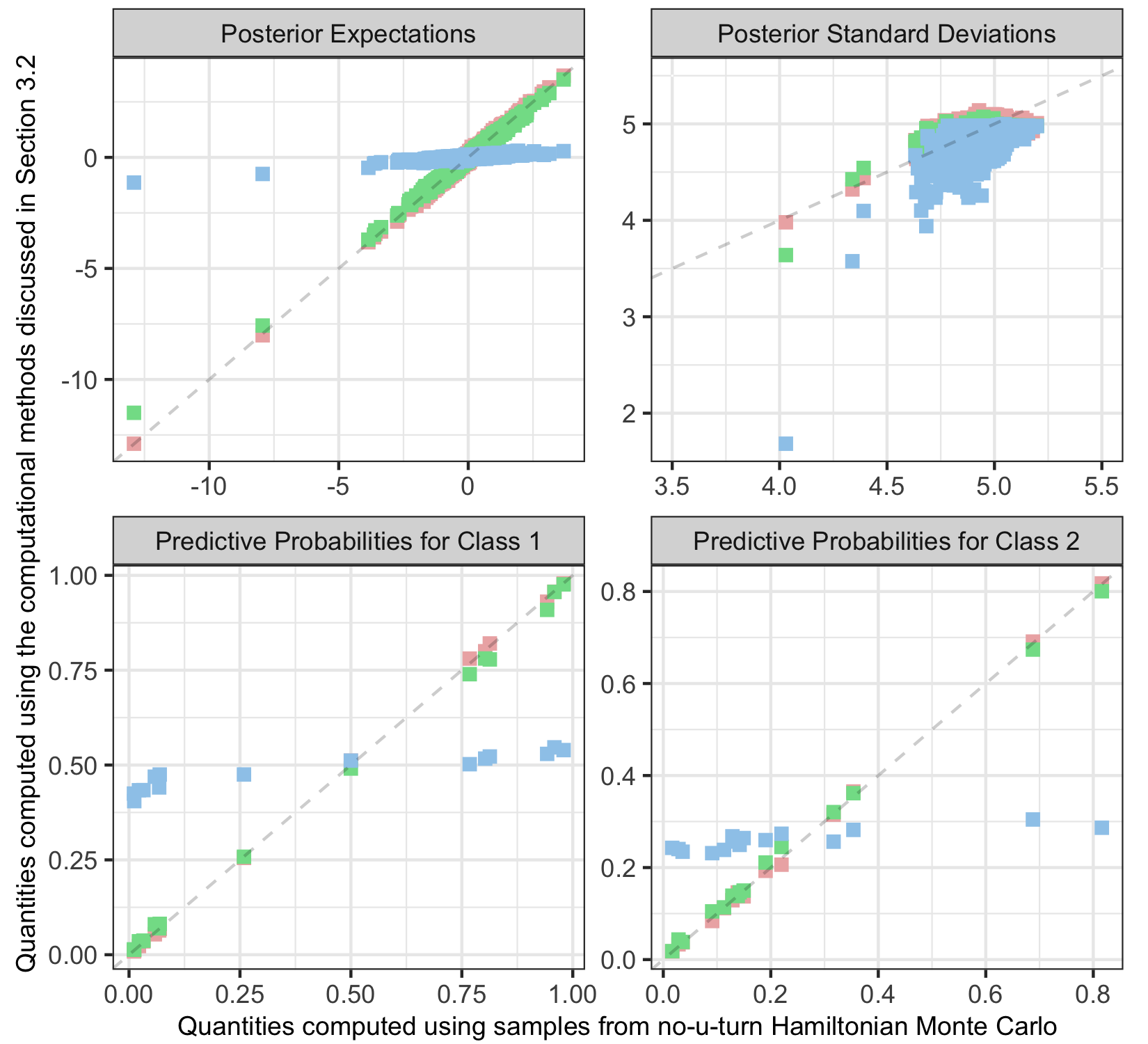}
	\caption{Comparison between the estimates of key functionals obtained under the  methods discussed in Sections \ref{sec.3.2.1} and \ref{sec.3.2.2} ($y$--axis), against those provided by the \texttt{rstan} implementation of the Hamiltonian no-u-turn sampler ($x$--axis). Red squares refer to Monte Carlo estimates based on i.i.d.\ samples from the exact \textsc{sun} posterior produced by Algorithm \ref{algo:samplingIID}, whereas blue and green squares denote the estimates provided by classical mean-field variational Bayes and by the proposed blocked partially-factorized  approximation, respectively.}
	\label{fig:F1}
\end{figure}

Figure \ref{fig:F1} compares the  Monte Carlo estimates for selected functionals of interest based on $5000$ \textsc{mcmc} samples from the Hamiltonian no-u-turn sampler (\texttt{R} package \texttt{rstan}), against those provided by the Monte Carlo and approximate methods discussed in Sections \ref{sec.3.2.1}--\ref{sec.3.2.2}. In particular, we compute such functionals using both $5000$ i.i.d.\ samples from the exact \textsc{sun} posterior provided by Algorithm \ref{algo:samplingIID}, and also by leveraging the  strategies associated with the blocked  partially-factorized  variational approximation  in Algorithm \ref{algo:2}. In computing such an approximation under the  sequential discrete choice multinomial probit model, we follow the guidelines in Remark \ref{remark1} and group those augmented data corresponding to the same unit $i$. We shall emphasize that when the coefficients are not shared across labels and have independent priors, the overlap among rows of $\bar{\bX}$ referring to the same unit $i$ is absent in sequential discrete choice representations. Hence, in this very specific case, we have that $\mbox{min}_{q(\bbeta,\bar{\bz}) \in \mathcal{Q}_{\textsc{pmf-b}}}\textsc{kl}[q(\bbeta, \bar{\bz})||p(\bbeta, \bar{\bz} \mid \by, \bX)]= \mbox{min}_{q(\bbeta,\bar{\bz}) \in \mathcal{Q}_{\textsc{pfm}}}\textsc{kl}[q(\bbeta, \bar{\bz})||p(\bbeta, \bar{\bz} \mid \by, \bX)]$. As we will discuss in the following, this blocking approach is more crucial for the multinomial probit models in Sections  \ref{sec.2.1}--\ref{sec.2.2}. To highlight the benefits of the  blocked partially-factorized approximation, we also compare results with classical mean-field variational Bayes enforcing independence between $\bbeta$ and $\bar{\bz}$ \citep{consonni_2007,girolami2006}.

As highlighted in Figure  \ref{fig:F1}, the two sampling-based methods provide comparable results in terms of inference and prediction. However, Algorithm \ref{algo:samplingIID} produces almost $75$ samples of $\bbeta$ per second, whereas the Hamiltonian no-u-turn sampler can only draw one sample every $3$ seconds. This massive computational cost makes state-of-the-art \textsc{mcmc} methods rapidly unfeasible in  large $p$ settings. We shall  highlight that by relying on i.i.d.\ samples, Algorithm \ref{algo:samplingIID} has also the advantage of avoiding the need of burn-in periods and convergence checks. However, as discussed in Sections \ref{sec.3.2} and \ref{sec.4}, Algorithm \ref{algo:samplingIID} scales poorly with sample size and, hence, it becomes impractical in studies with $n$ larger than a few hundreds. This motivates the blocked partially-factorized  approximation in Section \ref{sec.3.2.2}, that notably matches almost perfectly the Monte Carlo estimates in  such a high-dimensional setting  (see Figure  \ref{fig:F1}), and requires only $0.25$ seconds to converge and $16$ seconds to compute the different functionals. Classical mean-field variational Bayes has comparable running times, but the independence assumption between $\bbeta$ and $\bar{\bz}$ induces notable overshrinkage of both the locations and scales, which massively affects the estimation of the predictive probabilities. These results confirm and further clarify the findings in Section \ref{sec.4}.

\begin{figure}[t]
	\centering
		\includegraphics[scale=.167]{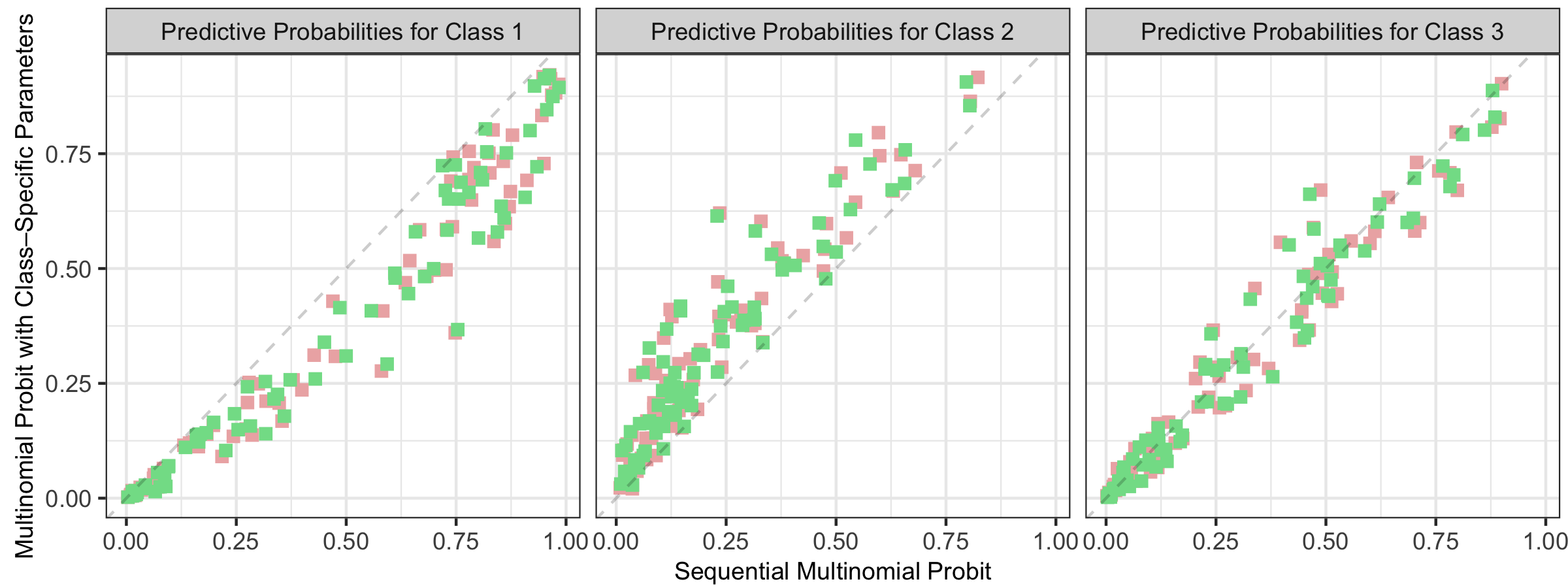}
		\caption{Comparison between the predictive probabilities provided by model \eqref{eq3} ($y$--axis) and  \eqref{eq5} ($x$--axis). The dataset has been divided into $6$ folds, and for each fold the predictive probabilities are computed using all the other available data as training set. Red squares refer to Monte Carlo estimates based on i.i.d.\ samples from the  \textsc{sun} posterior produced by Algorithm \ref{algo:samplingIID}, whereas the green squares denote the estimates provided by the proposed blocked  partially-factorized approximation.}
\label{fig:F2}
\end{figure}

Before concluding our analysis, we also implement the multinomial probit model with class-specific parameters presented in Section \ref{sec.2.2}, assuming independent standard normal errors. Due to the form of the dataset, the classical discrete multinomial probit  in Section \ref{sec.2.1} is not appropriate, since it would require a vector of covariates for each combination of  unit $i$ and lesion $l$, which is not the case for this study. Nonetheless, according to the results in Sections \ref{sec.2.1}, \ref{sec.2.2}, and \ref{sec.3}, models \eqref{eq1} and \eqref{eq3} induce posteriors with comparable dimensions and, hence, the performance of the multinomial probit with class-specific coefficients is also indicative of the one associated with the classical specification outlined in Section \ref{sec.2.1}. Here, we focus on comparing the computational and predictive performance between the already-implemented sequential formulation in \eqref{eq5} and the one having class-specific coefficients in \eqref{eq3}, considering the Monte Carlo and variational estimates discussed in Section \ref{sec.3.2}. Under model  \eqref{eq3}, blocking across units $i$ was more crucial to obtain accurate variational inference.  The Hamiltonian no-u-turn sampler faced, instead, severe mixing and convergence issues under model \eqref{eq3}, further highlighting major issues of \textsc{mcmc} in such settings.

Figure \ref{fig:F2} compares variational and Monte Carlo estimates of the predictive probabilities for all the units, under the two models. To estimate the predictive probabilities we split the dataset in six folds, four having $13$ observations and two having $12$ observations. Then, we compute the predictive probabilities for the observations in each fold, using the units in the remaining five folds to obtain the posterior distribution. As clarified in Figure \ref{fig:F2}, the two models provide similar, but not identical, predictive probabilities, whose values are almost the same when comparing the Monte Carlo and variational estimates. This result  confirms the excellent performance of the proposed  blocked partially-factorized approximation in high-dimensional settings, especially when the dimension of $\bbeta$ is higher than the sample size. Indeed, by slightly increasing the dimension of the training set, the number of $\bbeta$ samples per second produced by Algorithm \ref{algo:samplingIID} rapidly decreases from $75$ to $50$  in model~\eqref{eq5}, whereas the variational strategy still requires about $0.25$ seconds to converge and $16$ seconds to compute the functionals. The overall out-of-sample predictive accuracy under the two models is about $66.5\%$. Considering the simplicity of the multinomial probit models implemented, these values are quite satisfactory when compared with the $73.68\%$ accuracy obtained under sophisticated black-box machine learning algorithms \citep{mesejo2016}. 

\section{Discussion}
\label{sec.6}
This article provides novel conjugacy results and computational methods for a general class of multinomial probit models \citep{hausman1978,stern_1992,tutz1991} with Gaussian priors, and extends such properties to the entire class of \textsc{sun} \citep{arellano_2006} priors.  As discussed in Sections \ref{sec.3},  \ref{sec.4} and \ref{sec.5}, the availability of a  \textsc{sun} posterior allows major advances in terms of closed-form, Monte Carlo and approximate variational inference which cover a still unaddressed gap of \textsc{mcmc} methods in high-dimensional  studies. These settings are common  in a variety of fields, such as in medical applications collecting a huge number of predictors via state-of-the-art imaging technologies.

Our results open also several avenues for future research. For example, although Bayesian estimation and inference for the covariance matrix $\bSigma$ goes beyond the scope of this article, as mentioned in Sections \ref{sec.1} and \ref{sec.3.1}, the availability of a closed-form expression for the marginal likelihood in Corollary \ref{cor2} motivates promising advances in point estimation and full Bayesian inference also on  $\bSigma$, which deserve further exploration. The results in this article can be also included in more complex formulations. For instance, the sequential probit in~\eqref{eq5} has been used within Bayesian nonparametric  models for density regression based on probit stick-breaking process \citep{rodrig_2011}. Our findings could be useful in such settings to improve the computational performance and the theoretical treatment of predictor-dependent Bayesian nonparametric mixture models. Also extensions of our results to classification via Gaussian processes \citep{ras_2006,girolami2006,cao2020scalable} and state-space models \citep{fasano2021closed} are straightforward.  Finally, it would be also interesting to exploit the strategies in \citet{genton2018}, \citet{cao2019} and \citet{cao2021exploiting} to identify suitable blocks of augmented data in a more data-driven way, which can be applied to perform accurate variational inference not only in multinomial but also in binary  probit regression. Similarly, exploring other strategies for sampling from multivariate truncated normals, such  as the sequential Monte Carlo method in \citet{moffa2014sequential}, could further increase the impact of our findings.

% Acknowledgements should go at the end, before appendices and references

% Manual newpage inserted to improve layout of sample file - not
% needed in general before appendices/bibliography.

\appendix
\section*{Appendix A. Proofs}
\label{append}

{\bf Proof of Theorem  \ref{teo1}}. To prove Theorem \ref{teo1}, it suffices to first apply the Bayes rule and then recognize a \textsc{sun} density in the kernel of $p(\bbeta \mid \by, \bX)$. In particular, note that $p(\bbeta \mid \by, \bX) \propto p(\bbeta)p(\by \mid \bbeta, \bX) \propto \phi_q(\bbeta-\bxi;\bOmega)\Phi_h(\bgamma+\bDelta^{\intercal}\bar{\bOmega}{\phantom{.}}^{-1}\bomega^{-1}(\bbeta-\bxi);\bGamma-\bDelta^{\intercal}\bar{\bOmega}{\phantom{.}}^{-1}\bDelta)\Phi_m(\bar{\bX} \bbeta; \bLambda)$, and re-write $\Phi_m(\bar{\bX} \bbeta; \bLambda)$ as $\Phi_m(\bs^{-1} \bar{\bX} \bxi+(\bar{\bOmega}\bomega \bar{\bX}^{\intercal}\bs^{-1})^{\intercal}\bar{\bOmega}{\phantom{.}}^{-1}\bomega^{-1}(\bbeta-\bxi); \bs^{-1}(\bar{\bX}\bOmega\bar{\bX}^{\intercal}+\bLambda)\bs^{-1}-(\bar{\bOmega}\bomega \bar{\bX}^{\intercal}\bs^{-1})^{\intercal}\bar{\bOmega}{\phantom{.}}^{-1}\bar{\bOmega}\bomega\bar{\bX}^{\intercal}\bs^{-1})$. Replacing this quantity inside the kernel of the posterior and recalling proof of Corollary 4 in  \citet{Durante2018}, it follows that 
\begin{eqnarray*}
&&\Phi_h(\bgamma+\bDelta^{\intercal}\bar{\bOmega}\phantom{.}^{-1}\bomega^{-1}(\bbeta-\bxi);\bGamma-\bDelta^{\intercal}\bar{\bOmega}\phantom{.}^{-1}\bDelta)\Phi_m(\bar{\bX} \bbeta; \bLambda)\\
&&\quad =\Phi_{h+m}(\bgamma_{\post}{+}\bDelta_{\post}^{\intercal}\bar{\bOmega}_{\post}^{-1}\bomega_{\post}^{-1}(\bbeta-\bxi_{\post});\bGamma_{\post}{-}\bDelta_{\post}^{\intercal}\bar{\bOmega}_{\post}^{-1}\bDelta_{\post}),
\end{eqnarray*}
with $\bxi_{\post}$, $\bOmega_{\post}$, $\bDelta_{\post}$, $\bgamma_{\post}$ and $\bGamma_{\post}$ as in Theorem \ref{teo1}. Leveraging this equality and recalling that $\bxi_{\post}=\bxi$, $\bOmega_{\post}=\bOmega$, it can be immediately noticed that $p(\bbeta)p(\by \mid \bbeta, \bX) $ coincides with the kernel of the $\textsc{sun}$ in \eqref{eq9}, thereby proving Theorem \ref{teo1}. To prove that $\bOmega_{\post}^*$  is a correlation matrix it suffices to replace $\bI_n$ with $\bLambda$ in the proof of Corollary 4 in \citet{Durante2018}. \hfill\BlackBox

\vspace{10pt}

\noindent {\bf Proof of Corollary  \ref{cor2}.}
To show that $p(\by \mid \bX)$ can be written as in \eqref{eq10}, note that from the proof of Theorem \ref{teo1}, $p(\by,\bbeta \mid \bX)=p(\bbeta)\Phi_m(\bar{\bX} \bbeta; \bLambda)=p(\bbeta \mid \by,\bX)\Phi_{h+m}(\bgamma_{\post};\bGamma_{\post})/\Phi_h(\bgamma;\bGamma)$. Hence, $p(\by \mid \bX)=p(\by,\bbeta \mid \bX)/p(\bbeta \mid \by,\bX)=[p(\bbeta \mid \by,\bX)\Phi_{h+m}(\bgamma_{\post};\bGamma_{\post})/\Phi_h(\bgamma;\bGamma)]/p(\bbeta \mid \by,\bX)$, which implies $p(\by \mid \bX)=\Phi_{h+m}(\bgamma_{\post};\bGamma_{\post})/\Phi_h(\bgamma;\bGamma)$. \hfill\BlackBox

\vspace{10pt}

\noindent  {\bf Proof of Corollary  \ref{cor3}.}
To prove Corollary \ref{cor3} simply notice that  \eqref{eq11} is the ratio between the marginal likelihoods of the new expanded dataset and the original one (i.e., the one without the additional unit with response  $y_{\new}=l$ and covariates $\bx_{\new}$). Hence, the expression for the predictive probabilities follows  from Corollary  \ref{cor2} after noting that, due to the conditional independence assumption in   \eqref{eq1}, \eqref{eq3} or \eqref{eq5}, $p(\by \mid \bX,\bx_{\new})=p(\by \mid \bX)$. \hfill\BlackBox


\vskip 0.2in
\begin{thebibliography}{61}
\providecommand{\natexlab}[1]{#1}
\providecommand{\url}[1]{\texttt{#1}}
\expandafter\ifx\csname urlstyle\endcsname\relax
  \providecommand{\doi}[1]{doi: #1}\else
  \providecommand{\doi}{doi: \begingroup \urlstyle{rm}\Url}\fi

\bibitem[Agarwal et~al.(2014)Agarwal, Ranjan, and Chipman]{agarwal2014}
R.~Agarwal, P.~Ranjan, and H.~Chipman.
\newblock A new {B}ayesian ensemble of trees approach for land cover
  classification of satellite imagery.
\newblock \emph{Canadian Journal of Remote Sensing}, 39:\penalty0 507--520,
  2014.

\bibitem[Agresti(2013)]{agrest_2013}
A.~Agresti.
\newblock \emph{Categorical Data Analysis (Third Edition)}.
\newblock Wiley, 2013.

\bibitem[Albert and Chib(1993)]{Albert_1993}
J.H. Albert and S.~Chib.
\newblock Bayesian analysis of binary and polychotomous response data.
\newblock \emph{Journal of the American Statistical Association}, 88:\penalty0
  669--679, 1993.

\bibitem[Albert and Chib(2001)]{albert2001}
J.H. Albert and S.~Chib.
\newblock Sequential ordinal modeling with applications to survival data.
\newblock \emph{Biometrics}, 57:\penalty0 829--836, 2001.

\bibitem[Arellano-Valle and Azzalini(2006)]{arellano_2006}
R.B. Arellano-Valle and A.~Azzalini.
\newblock On the unification of families of skew-normal distributions.
\newblock \emph{Scandinavian Journal of Statistics}, 33:\penalty0 561--574,
  2006.

\bibitem[Azzalini(1985)]{azza_1985}
A.~Azzalini.
\newblock A class of distributions which includes the normal ones.
\newblock \emph{Scandinavian Journal of Statistics}, 12:\penalty0 171--178,
  1985.

\bibitem[Azzalini and Bacchieri(2010)]{azzalini_2010}
A.~Azzalini and A.~Bacchieri.
\newblock A prospective combination of phase {II} and phase {III} in drug
  development.
\newblock \emph{Metron}, 68:\penalty0 347--369, 2010.

\bibitem[Azzalini and Capitanio(2013)]{azzalini_2013}
A.~Azzalini and A.~Capitanio.
\newblock \emph{{The Skew-Normal and Related Families}}.
\newblock Cambridge University Press, 2013.

\bibitem[Azzalini and Dalla~Valle(1996)]{azza_1996}
A.~Azzalini and A.~Dalla~Valle.
\newblock The multivariate skew-normal distribution.
\newblock \emph{Biometrika}, 83:\penalty0 715--726, 1996.

\bibitem[Bishop(2006)]{bishop2006}
C.M. Bishop.
\newblock \emph{Pattern {R}ecognition and {M}achine {L}earning}.
\newblock Springer, 2006.

\bibitem[Blei et~al.(2017)Blei, Kucukelbir, and McAuliffe]{blei2017}
D.M. Blei, A.~Kucukelbir, and J.D. McAuliffe.
\newblock Variational inference: {A} review for statisticians.
\newblock \emph{Journal of the American Statistical Association}, 112:\penalty0
  859--877, 2017.

\bibitem[B{\"o}rsch-Supan and Hajivassiliou(1993)]{borsch_1993}
A.~B{\"o}rsch-Supan and V.A. Hajivassiliou.
\newblock Smooth unbiased multivariate probability simulators for maximum
  likelihood estimation of limited dependent variable models.
\newblock \emph{Journal of Econometrics}, 58:\penalty0 347--368, 1993.

\bibitem[Botev(2017)]{botev_2017}
Z.I. Botev.
\newblock The normal law under linear restrictions: {S}imulation and estimation
  via minimax tilting.
\newblock \emph{Journal of the Royal Statistical Society: Series B},
  79:\penalty0 125--148, 2017.

\bibitem[Burgette and Nordheim(2012)]{burgette2012}
L.F. Burgette and E.V. Nordheim.
\newblock {The trace restriction: An alternative identification strategy for
  the Bayesian multinomial probit model}.
\newblock \emph{Journal of Business \& Economic Statistics}, 30:\penalty0
  404--410, 2012.

\bibitem[Cao et~al.(2019)Cao, Genton, Keyes, and Turkiyyah]{cao2019}
J.~Cao, M.G. Genton, D.E. Keyes, and G.M. Turkiyyah.
\newblock Hierarchical--block conditioning approximations for high-dimensional
  multivariate normal probabilities.
\newblock \emph{Statistics and Computing}, 29:\penalty0 585--598, 2019.

\bibitem[Cao et~al.(2021)Cao, Genton, Keyes, and Turkiyyah]{cao2021exploiting}
J.~Cao, M.G. Genton, D.E. Keyes, and G.M. Turkiyyah.
\newblock Exploiting low-rank covariance structures for computing
  high-dimensional normal and student-t probabilities.
\newblock \emph{Statistics and Computing}, 31:\penalty0 2, 2021.

\bibitem[Cao et~al.(2022)Cao, Durante, and Genton]{cao2020scalable}
J.~Cao, D.~Durante, and M.G. Genton.
\newblock Scalable computation of predictive probabilities in probit models
  with {G}aussian process priors.
\newblock \emph{Journal of Computational and Graphical Statistics}, (forthcoming),
  2022.

\bibitem[Chan and Jeliazkov(2009)]{chan2009}
J.~C.-C. Chan and I.~Jeliazkov.
\newblock {MCMC} estimation of restricted covariance matrices.
\newblock \emph{Journal of Computational and Graphical Statistics},
  18:\penalty0 457--480, 2009.

\bibitem[Chen and Kuo(2002)]{chen2002}
Z.~Chen and L.~Kuo.
\newblock Discrete choice models based on the scale mixture of multivariate
  normal distributions.
\newblock \emph{Sankhy{\=a}, Series B}, 64:\penalty0 192--213, 2002.

\bibitem[Chopin(2011)]{chop_2011}
N.~Chopin.
\newblock Fast simulation of truncated {G}aussian distributions.
\newblock \emph{Statistics and Computing}, 21:\penalty0 275--288, 2011.

\bibitem[Chopin and Ridgway(2017)]{chopin_2017}
N.~Chopin and J.~Ridgway.
\newblock Leave {P}ima indians alone: {B}inary regression as a benchmark for
  {B}ayesian computation.
\newblock \emph{Statistical Science}, 32:\penalty0 64--87, 2017.

\bibitem[Consonni and Marin(2007)]{consonni_2007}
G.~Consonni and J.M. Marin.
\newblock Mean-field variational approximate {B}ayesian inference for latent
  variable models.
\newblock \emph{Computational Statistics \& Data Analysis}, 52:\penalty0
  790--798, 2007.

\bibitem[Daganzo(1979)]{daganzo2014}
C.~Daganzo.
\newblock \emph{Multinomial Probit}.
\newblock Academic Press, 1979.

\bibitem[Dow and Endersby(2004)]{dow2004}
J.K. Dow and J.W. Endersby.
\newblock Multinomial probit and multinomial logit: a comparison of choice
  models for voting research.
\newblock \emph{Electoral Studies}, 23:\penalty0 107--122, 2004.

\bibitem[Durante(2019)]{Durante2018}
D.~Durante.
\newblock Conjugate {B}ayes for probit regression via unified skew-normal
  distributions.
\newblock \emph{Biometrika}, 106:\penalty0 765--779, 2019.

\bibitem[Fasano et~al.(2021)Fasano, Rebaudo, Durante, and
  Petrone]{fasano2021closed}
A.~Fasano, G.~Rebaudo, D.~Durante, and S.~Petrone.
\newblock A closed-form filter for binary time series.
\newblock \emph{Statistics and Computing}, 31:\penalty0 47, 2021.

\bibitem[Fasano et~al.(2022)Fasano, Durante, and
  Zanella]{fasano2019asymptotically}
A.~Fasano, D.~Durante, and G.~Zanella.
\newblock Scalable and accurate variational {B}ayes for high--dimensional
  binary regression models.
\newblock \emph{Biometrika}, (forthcoming), 2022.

\bibitem[Gelman et~al.(2008)Gelman, Jakulin, Pittau, and Su]{gelman2008}
A.~Gelman, A.~Jakulin, M.G. Pittau, and Y.S. Su.
\newblock A weakly informative default prior distribution for logistic and
  other regression models.
\newblock \emph{The Annals of Applied Statistics}, 2:\penalty0 1360--1383,
  2008.

\bibitem[Genton et~al.(2018)Genton, Keyes, and Turkiyyah]{genton2018}
M.G. Genton, D.E. Keyes, and G.M. Turkiyyah.
\newblock Hierarchical decompositions for the computation of high-dimensional
  multivariate normal probabilities.
\newblock \emph{Journal of Computational and Graphical Statistics},
  27:\penalty0 268--277, 2018.

\bibitem[Genz(1992)]{genz_1992}
A.~Genz.
\newblock Numerical computation of multivariate normal probabilities.
\newblock \emph{Journal of Computational and Graphical Statistics}, 1:\penalty0
  141--149, 1992.

\bibitem[Geweke et~al.(1994)Geweke, Keane, and Runkle]{geweke1994}
J.~Geweke, M.~Keane, and D.~Runkle.
\newblock Alternative computational approaches to inference in the multinomial
  probit model.
\newblock \emph{The Review of Economics and Statistics}, 76:\penalty0 609--632,
  1994.

\bibitem[Girolami and Rogers(2006)]{girolami2006}
M.~Girolami and S.~Rogers.
\newblock Variational {B}ayesian multinomial probit regression with {G}aussian
  process priors.
\newblock \emph{Neural Computation}, 18:\penalty0 1790--1817, 2006.

\bibitem[Girolami and Zhong(2007)]{girolami2007}
M.~Girolami and M.~Zhong.
\newblock Data integration for classification problems employing {G}aussian
  process priors.
\newblock In \emph{Advances in Neural Information Processing Systems},
  volume~20, pages 465--472, 2007.

\bibitem[Greene(2003)]{greene2003}
W.H. Greene.
\newblock \emph{Econometric Analysis}.
\newblock Prentice Hall, 2003.

\bibitem[Gupta et~al.(2013)Gupta, Aziz, and Ning]{gupta_2013}
A.K. Gupta, M.A. Aziz, and W.~Ning.
\newblock On some properties of the unified skew-normal distribution.
\newblock \emph{Journal of Statistical Theory and Practice}, 7:\penalty0
  480--495, 2013.

\bibitem[Hausman and Wise(1978)]{hausman1978}
J.A. Hausman and D.A. Wise.
\newblock A conditional probit model for qualitative choice: {D}iscrete
  decisions recognizing interdependence and heterogeneous preferences.
\newblock \emph{Econometrica: Journal of the Econometric Society}, 46:\penalty0
  403--426, 1978.

\bibitem[Hoffman and Gelman(2014)]{hoff_2014}
M.D. Hoffman and A.~Gelman.
\newblock The {N}o-{U}-turn sampler: {A}daptively setting path lengths in
  {H}amiltonian {M}onte {C}arlo.
\newblock \emph{Journal of Machine Learning Research}, 15:\penalty0 1593--1623,
  2014.

\bibitem[Holmes and Held(2006)]{holmes_2006}
C.C. Holmes and L.~Held.
\newblock Bayesian auxiliary variable models for binary and multinomial
  regression.
\newblock \emph{Bayesian Analysis}, 1:\penalty0 145--168, 2006.

\bibitem[Horrace(2005)]{horrace2005}
W.C. Horrace.
\newblock Some results on the multivariate truncated normal distribution.
\newblock \emph{Journal of Multivariate Analysis}, 94:\penalty0 209--221, 2005.

\bibitem[Imai and Van~Dyk(2005)]{imai2005}
K.~Imai and D.A. Van~Dyk.
\newblock A {B}ayesian analysis of the multinomial probit model using marginal
  data augmentation.
\newblock \emph{Journal of Econometrics}, 124:\penalty0 311--334, 2005.

\bibitem[Johndrow et~al.(2013)Johndrow, Dunson, and Lum]{johndrow2013}
J.E. Johndrow, D.B. Dunson, and K.~Lum.
\newblock Diagonal orthant multinomial probit models.
\newblock In \emph{Artificial Intelligence and Statistics},  volume~31, pages 29--38, 2013.

\bibitem[Johndrow et~al.(2019)Johndrow, Smith, Pillai, and
  Dunson]{Johndrow2018}
J.E. Johndrow, A.~Smith, N.~Pillai, and D.B. Dunson.
\newblock {MCMC} for imbalanced categorical data.
\newblock \emph{Journal of the American Statistical Association}, 114:\penalty0
  1394--1403, 2019.

\bibitem[Kindo et~al.(2016)Kindo, Wang, and Pe{\~n}a]{kindo2016}
B.P. Kindo, H.~Wang, and E.A. Pe{\~n}a.
\newblock Multinomial probit {B}ayesian additive regression trees.
\newblock \emph{Stat}, 5:\penalty0 119--131, 2016.

\bibitem[Knowles and Minka(2011)]{knowles2011}
D.A. Knowles and T.~Minka.
\newblock Non-conjugate variational message passing for multinomial and binary
  regression.
\newblock In \emph{Advances in Neural Information Processing Systems},
  volume~24, pages 1701--1709, 2011.

\bibitem[Kullback and Leibler(1951)]{kullback_1951}
S.~Kullback and R.A. Leibler.
\newblock On information and sufficiency.
\newblock \emph{The Annals of Mathematical Statistics}, 22:\penalty0 79--86,
  1951.

\bibitem[Maddala(1986)]{maddala1986}
G.S. Maddala.
\newblock \emph{{Limited-Dependent and Qualitative Variables in Econometrics}}.
\newblock Cambridge University Press, 1986.

\bibitem[McCulloch and Rossi(1994)]{mcculloch1994}
R.E. McCulloch and P.E. Rossi.
\newblock An exact likelihood analysis of the multinomial probit model.
\newblock \emph{Journal of Econometrics}, 64:\penalty0 207--240, 1994.

\bibitem[McCulloch et~al.(2000)McCulloch, Polson, and Rossi]{mcculloch2000}
R.E. McCulloch, N.G. Polson, and P.E. Rossi.
\newblock A {B}ayesian analysis of the multinomial probit model with fully
  identified parameters.
\newblock \emph{Journal of Econometrics}, 99:\penalty0 173--193, 2000.

\bibitem[McFadden(1989)]{mcfadden_1989}
D.~McFadden.
\newblock A method of simulated moments for estimation of discrete response
  models without numerical integration.
\newblock \emph{Econometrica: Journal of the Econometric Society}, 57:\penalty0
  995--1026, 1989.

\bibitem[Mesejo et~al.(2016)Mesejo, Pizarro, Abergel, Rouquette, Beorchia,
  Poincloux, and Bartoli]{mesejo2016}
P.~Mesejo, D.~Pizarro, A.~Abergel, O.~Rouquette, S.~Beorchia, L.~Poincloux, and
  A.~Bartoli.
\newblock Computer-aided classification of gastrointestinal lesions in regular
  colonoscopy.
\newblock \emph{IEEE Transactions on Medical Imaging}, 35:\penalty0 2051--2063,
  2016.

\bibitem[Moffa and Kuipers(2014)]{moffa2014sequential}
G.~Moffa and J.~Kuipers.
\newblock {Sequential Monte Carlo EM for multivariate probit models}.
\newblock \emph{Computational Statistics \& Data Analysis}, 72:\penalty0
  252--272, 2014.

\bibitem[Natarajan et~al.(2000)Natarajan, McCulloch, and
  Kiefer]{natarajan2000monte}
R.~Natarajan, C.E. McCulloch, and N.M. Kiefer.
\newblock {A Monte Carlo EM method for estimating multinomial probit models}.
\newblock \emph{Computational Statistics \& Data Analysis}, 34:\penalty0
  33--50, 2000.

\bibitem[Nobile(1998)]{nobile1998}
A.~Nobile.
\newblock {A hybrid Markov chain for the {B}ayesian analysis of the multinomial
  probit model}.
\newblock \emph{Statistics and Computing}, 8:\penalty0 229--242, 1998.

\bibitem[Park and Van~Dyk(2009)]{park2009}
T.~Park and D.A. Van~Dyk.
\newblock Partially collapsed {G}ibbs samplers: Illustrations and applications.
\newblock \emph{Journal of Computational and Graphical Statistics},
  18:\penalty0 283--305, 2009.

\bibitem[Rasmussen and Williams(2006)]{ras_2006}
C.E. Rasmussen and C.K. Williams.
\newblock \emph{{Gaussian Processes for Machine Learning}}.
\newblock MIT Press, 2006.

\bibitem[Riihim{\"a}ki et~al.(2013)Riihim{\"a}ki, Jyl{\"a}nki, and
  Vehtari]{riihimaki2013}
J.~Riihim{\"a}ki, P.~Jyl{\"a}nki, and A.~Vehtari.
\newblock Nested expectation propagation for {G}aussian process classification
  with a multinomial probit likelihood.
\newblock \emph{Journal of Machine Learning Research}, 14:\penalty0 75--109,
  2013.

\bibitem[Rodriguez and Dunson(2011)]{rodrig_2011}
A.~Rodriguez and D.B. Dunson.
\newblock Nonparametric {B}ayesian models through probit stick-breaking
  processes.
\newblock \emph{Bayesian Analysis}, 6:\penalty0 145--178, 2011.

\bibitem[Rogers and Girolami(2007)]{rogers2007}
S.~Rogers and M.~Girolami.
\newblock Multi-class semi-supervised learning with the $\epsilon$-truncated
  multinomial probit {G}aussian process.
\newblock In \emph{Journal of Machine Learning Research, Workshop \&
  Proceedings}, volume~1, pages 17--32, 2007.

\bibitem[Stern(1992)]{stern_1992}
S.~Stern.
\newblock A method for smoothing simulated moments of discrete probabilities in
  multinomial probit models.
\newblock \emph{Econometrica: Journal of the Econometric Society}, 60:\penalty0
  943--952, 1992.

\bibitem[Tutz(1991)]{tutz1991}
G.~Tutz.
\newblock Sequential models in categorical regression.
\newblock \emph{Computational Statistics \& Data Analysis}, 11:\penalty0
  275--295, 1991.

\bibitem[Zhang et~al.(2006)Zhang, Boscardin, and Belin]{zhang2006}
X.~Zhang, W.J. Boscardin, and T.R. Belin.
\newblock Sampling correlation matrices in {B}ayesian models with correlated
  latent variables.
\newblock \emph{Journal of Computational and Graphical Statistics},
  15:\penalty0 880--896, 2006.

\end{thebibliography}
\end{document}